\documentclass[12pt,twoside]{article}   

\usepackage[super,sort&compress,comma]{natbib}

\usepackage[dvipsnames]{xcolor}
\usepackage{mathtools, nccmath,commath}

\usepackage{fancyhdr}		

\def\SPSB#1#2{\rlap{\textsuperscript{{#1}}}\SB{#2}}

\def\SB#1{\textsubscript{{#1}}}

\newcommand{\argmin}{\arg\!\min}

\usepackage[hyphens]{url}
\usepackage{hyperref}
\usepackage[hyphenbreaks]{breakurl}
\usepackage{soul}
\usepackage[section]{placeins}   %
\usepackage{siunitx}
\usepackage{graphicx}
\usepackage{amsmath}
\usepackage{tikz}
\usetikzlibrary{spy}
\makeatletter \renewcommand\@biblabel[1]{$^{#1}$} \makeatother
\DeclareMathOperator{\arctantwo}{arctan2}
\newcommand{\cen}[1]{\begin{center} #1 \end{center}}

 \usepackage[pagewise,mathlines,edtable]{lineno}

\usepackage{amssymb}

\hypersetup{ colorlinks,
    citecolor=blue,
    filecolor=blue,
    linkcolor=blue,
    urlcolor=blue
}

\usepackage{textcomp}
\usepackage{subcaption}
\usepackage{wrapfig}

\usepackage{xcolor}

\definecolor{gray}{rgb}{0.6,0.6,0.6}
\definecolor{red}{rgb}{0.85,0,0}
\definecolor{green}{rgb}{0,0.85,0}
\definecolor{blue}{rgb}{0,0,0.85}
\definecolor{beige}{rgb}{0.92,0.87,0.78}
\usepackage[all]{hypcap}    

\graphicspath{ {./} }

\begin{document}

\cen{\sf {\Large {\bfseries Physics-Based Iterative Reconstruction for Dual Source and Flying Focal Spot Computed Tomography} \\  
\vspace*{5mm}
Xiao Wang, \hspace{4mm}   Robert D. MacDougall} \\
Department of Radiology, Boston Children’s Hospital, Harvard Medical School, Boston, Massachusetts 02115 \vspace{3mm}\\
{\Large Peng Chen} \\
National Institute of Advanced Industrial Science and Technology, Japan \\
RIKEN Center for Computational Science, Japan
\vspace{3mm}\\
{\Large Charles A. Bouman} \\
School of Electrical and Computer Engineering, Purdue University, West Lafayette, Indiana 47907
\vspace{3mm}\\
{\Large Simon K. Warfield} \\
Department of Radiology, Boston Children’s Hospital,
Harvard Medical School, Boston, Massachusetts 02115 \\
\vspace{3mm}
}

\pagenumbering{roman}
\setcounter{page}{1}
\pagestyle{plain}

\begin{abstract}
\noindent \\{\bf Purpose:}
For single source helical Computed Tomography (CT), both Filtered-Back Projection (FBP) and statistical iterative reconstruction have been investigated. However for dual source CT with flying focal spot (DS-FFS CT), statistical iterative reconstruction that accurately models the scanner geometry and acquisition physics remains unknown to researchers. Therefore, our purpose is to present a novel physics-based iterative reconstruction method for DS-FFS CT and assess its image quality.\\
{\bf Methods:} Our algorithm uses precise physics models to reconstruct from the native cone-beam geometry and interleaved dual source helical trajectory of a DS-FFS CT. To do so, we construct a noise physics model to represent data acquisition noise and a prior image model to represent image noise and texture. In addition, we design forward system models to compute the locations of deflected focal spots, the dimension and sensitivity of voxels and detector units, as well as the length of intersection between X-rays and voxels. The forward system models further represent the coordinated movement between the dual sources by computing their X-ray coverage gaps and overlaps at an arbitrary helical pitch. With the above models, we reconstruct images by  using an advanced Consensus Equilibrium (CE) numerical method to compute the {\em maximum a posteriori} estimate to a joint optimization problem that simultaneously fits all models. \\
{\bf Results:} We compared our reconstruction with Siemens ADMIRE, which is the clinical standard hybrid iterative reconstruction (IR) method for DS-FFS CT, in terms of spatial resolution, noise profile and image artifacts through both phantoms and clinical scan datasets. Experiments show that our reconstruction has a higher spatial resolution, with a Task-Based Modulation Transfer Function (MTF\textsubscript{task}) consistently higher than the clinical standard hybrid IR. In addition, our reconstruction shows a reduced magnitude of image undersampling artifacts than the clinical standard.\\
{\bf Conclusions:} By modeling a precise geometry and avoiding data rebinning or interpolation, our physics-based reconstruction achieves a higher spatial resolution and fewer image artifacts with smaller magnitude than the clinical standard hybrid IR.

\noindent \\{\bf Keywords:} Dual Source CT, Flying Focal Spot, Statistical Iterative Reconstruction.
\end{abstract}

\setlength{\baselineskip}{0.7cm}      

\pagenumbering{arabic}
\setcounter{page}{1}
\pagestyle{fancy}

\section{Introduction}
\label{sec:intro}
Dual source computed tomography (CT) is a popular imaging modality that mounts two X-ray sources and detectors on the same rotating gantry, and uses a high helical pitch to rapidly acquire projections with high temporal resolution. With a dual source CT design, radiologists can examine heart and coronary arteries with much fewer motion artifacts than the single source design that has few detector rows~\cite{Petersilka08}. In addition, patients who have trouble holding still on a patient bed, such as children and patients with neurological disorders, are less likely to require anesthesia when being scanned by a dual source CT, thereby reducing the exam cost and sedation medical risks~\cite{Gottumukkala19}. 

A high helical pitch for dual source scanner, however, reduces sampling rate and in turn causes a degraded spatial resolution and more undersampling artifacts. To maintain spatial resolution and minimize artifacts, there are two approaches among CT vendors to increase sampling rate while keeping a high pitch. The first approach is to increase the CT detector resolution so that each projection has more samples. An example is the Aquilion Precision Scanner of Cannon with 0.25 mm detector resolution and a small field-of-view. The Aquilion Precision Scanner, however, is not in clinical use and its diagnostic accuracy is unknown. The second approach is to increase the total number of projections, such as with the widely used Siemens Somatom Force Scanner. Through the {\em flying focal spot} (FFS) technology, a dual source scanner takes multiple interleaved projections at each view angle without significantly lengthening the scan time. 

To reconstruct from the projections of a dual source flying focal spot CT, abbreviated as DS-FFS CT, Filtered Back-Projection (FBP) is the dominant approach for DS-FFS CT in the published works~\cite{Flohr05, Kyriakou06,Flohr08,Flohr09}. The general principle of FBP is to interpolate discretely sampled projections and transform them back to a reconstruction in the spatial domain through Fourier and inverse Fourier Transform~\cite{Turbell-Dissertation, Wang93}. As of today, there is no exact FBP implementation, such as the algorithms of Katsevich~\cite{Katsevich02}, that can fully model the geometry of DS-FFS CT and the existing methods for DS-FFS CT are all approximate FBP techniques that rebin the helical projections into parallel-beams for reconstruction~\cite{Noo99,Kachelriess00,Flohr05,Flohr08}. A clear advantage of the approximate FBP methods is that they enable a short reconstruction time by simplifying geometry and saving computations. The disadvantage is that these approximate FBP methods produce undesirable cone-beam artifacts when the cone angle is large and diagnostic errors may increase due to the cone-beam artifacts~\cite{Thibault07,Hsieh2013}. In addition, both the exact and the approximate FBP
are ``continuous to discrete" methods that map continuously sampled detector signals to discrete image voxels. The actual CT projections, however, are all discretely sampled instead of continuously sampled. Therefore, data interpolation is heavily used in the preprocessing steps of the FBP methods to transform discrete projections into continuous signals before performing a reconstruction. When the discrete projections are not sufficiently many, such as when the pitch is high, data interpolation may limit the achievable spatial resolution of the reconstruction and cause undersampling artifacts~\cite{Thibault07}. In summary, the advantages and disadvantages of FBP are listed on the left side of Table~\ref{table:algorithm-comparison}.

\begin{table}
{\small
\centering
\begin{tabular}{|c|c|c|}
\hline
Scanner & FBP &  Statistical IR \\
\hline
\begin{tabular}{@{}l@{}}Single Source \\ No FFS\end{tabular} & \begin{tabular}{@{}l@{}}  Fourier and inverse Fourier Transform \\ Interpolate projections \\May approximate geometry by rebinning  \\  Inferior quality at low dose \\ Artifacts at large cone angle \\ Popular for clinical practice  \\ Short reconstruction time in seconds \end{tabular} & \begin{tabular}{@{}l@{}}Linear Algebra and Bayesian \\ No interpolation \\ No rebinning  \\ Preserved quality at low dose \\ Fewer artifacts at large cone angle \\ Unpopular for clinical practice \\ Much longer runtime in hours \end{tabular} \\
\hline
\begin{tabular}{@{}l@{}}Dual Source \\ with FFS\end{tabular} &	\begin{tabular}{@{}l@{}}All above features\\ Always perform projection rebinning \\ Enable a higher sampling rate \\ than without FFS \\ Enable a much higher pitch \\ than Single Source\end{tabular} & No implementation\\
\hline
\end{tabular}
\caption{FBP and statistical IR algorithmic comparison.}
\label{table:algorithm-comparison}
 \vspace{1em}
 }
\end{table}

In contrast, iterative reconstruction (IR) formulates the final image as the solution to an optimization problem and solves the problem in an iterative fashion~\cite{Hsieh2013}. Among different IR methods, statistical IR reconstruction is a special one with the following five components~\cite{Fessler94}:
\begin{enumerate}
    \item {\bf An image model} that expresses the unknown object to be reconstructed in terms of voxels to be estimated from the projection data. In a Bayesian estimation framework, the image model is a prior model, such as L1, L2 norm and Total Variation.
    \item 	{\bf A system model} that represents radon transform, scanner geometry as well as acquisition physics. For tomography imaging, the system model can be expressed in the linear form $Y=AX+E$, where $Y$ is sinogram projections, $A$ is a system matrix, $X$ is the reconstructed image, and $E$ is the unknown measurement error.
    \item {\bf A statistical model} that describes how the noisy projection measurements vary around their ideal values and often the projection noise is assumed to have Gaussian or Poisson distribution. The statistical model also assigns weights to each projection based on the projection noise variance and penalizes noisy projections.
    \item {\bf A cost function} that is to be minimized to estimate the image voxels. 
    \item {\bf An iterative numerical algorithm} for minimizing the cost function.
\end{enumerate}
A unique advantage of statistical IR is that the system model enables the statistical IR to have a ``discrete-to-discrete" mapping from discretely sampled projections to discrete image voxels. Therefore, no data interpolation or rebinning is required and the algorithm operates directly on discrete measurements~\cite{XWang19,XWang16,XWang17-2}. In addition, statistical IR has great flexibility to incorporate precise scanner hardware characteristics into its data acquisition model~\cite{Zhang14,Thibault07}. Therefore, statistical IR has the potential to be more faithful to the true acquisition physics and the scanner geometry than FBP. Because of these distinct advantages, statistical IR often produces clearer image details with fewer artifacts than FBP, especially when the radiation doses are low and the number of projections is limited~\cite{Pickhardt12,Thibault07,Zhang14,Widmann15,Wang2006,XWang16}. The number of operations for statistical IR, however, is several magnitudes more than those for FBP~\cite{XWang16,Hsieh2013}. Therefore, statistical IR has a slow reconstruction time and is unacceptable for emergency medicine. In addition to the above disadvantages, the existing implementation for statistical IR is only applicable to single source CT and has no implementation for DS-FFS scanner. The commercial implementation for DS-FFS scanner, such as Siemens ADMIRE, is a hybrid algorithm that has some features of IR while keeping the core operations of weighted FBP. It has a forward model that performs weighted FBP operations and a prior model for image denoising and artifact removal. In addition, the forward model has extra loops for artifacts removal and statistical weighting~\cite{ADMIRE_white_paper}. Therefore, Siemens ADMIRE meets all the above conditions for statistical IR except point 2 because the forward model for Siemens ADMIRE does not have a system matrix and is not based upon the linear system model of the form $Y=AX+E$. In summary, the advantages and disadvantages of statistical IR are listed on the right side of Table~\ref{table:algorithm-comparison}~\cite{open_software_mbir, Fessler15}.

To address these limitations, this paper proposes a statistical IR algorithm, {\em The Joint Estimation for Native Geometry} (JENG), for DS-FFS CT scanner. This algorithm not only accounts for the true acquisition geometry and jointly estimates images from each X-ray source and focal spot, but also provides a detailed description on how to construct a linear forward system model for DS-FFS CT. In Sec.~\ref{subsec:FFS},
we propose a novel physics-based system model for JENG that imitates the flying focal spot data acquisition and native cone-beam geometry without projection interpolation, rebinning or completion.
Sec.~\ref{subsec:dual-source} characterizes the interleaved dual source helical trajectory at a high helical pitch. With precise knowledge of the scanner movement, we then reconstruct images by using Consensus Equilibrium to compute the {\em maximum a posteriori} estimate to a joint optimization problem that simultaneously fits projections from all focal spots and source-detector pairs. In Sec.~\ref{sec:results}, we evaluated image spatial resolution and artifacts of JENG on a standard ACR 464 phantom with respect to Task-Based Modulation Transfer Function (MTF\textsubscript{task}), Noise Power Spectrum (NPS) and undersampling artifacts. Experimental results show that JENG has fewer image artifacts and a much higher MTF\textsubscript{task} than the clinical standard hybrid IR method (Siemens ADMIRE). In addition, we also subjectively evaluated the spatial resolution and artifacts of JENG and the clinical standard method on 5 thoracic datasets and 3 abdominal datasets.

\section{Related Work}
\label{subsec:related}
Flohr and Kachelrie{\ss}'s papers analyze focal spot movement on a single source CT, and propose a rebinning FBP method to approximate interleaved helical multislice projection data as progressive-view and parallel-beam data~\cite{Flohr05, Kyriakou06}. After the rebinning, a 2D FBP is performed on the rebinned projection data, slice by slice. Such a method has three issues: (1) loss of spatial resolution from interpolation and geometry approximation, and the blurriness is often more pronounced in CT datasets with a high pitch~\cite{Zhao07}; (2) loss of spatial resolution on the edge of each image slice, as Flohr's reconstruction is a stack of 2D images rather than a fully 3D volume; and (3) noticeable aliasing artifacts in each image slice and windmill artifacts across image slices, especially when the CT cone angle is large~\cite{Thibault07,Hsieh2013}.

Flohr and his collaborators further extend the above work to DS-FFS CT~\cite{Flohr08,Flohr09}. As a dual source CT gantry often has limited space to fit two full size detectors, the dual source gantry often has a wide detector covering the full field of view and a narrow detector covering a truncated center view. Therefore, voxels that are outside of the narrow detector's truncated field-of-view receive no projections from the narrow detector, despite that these voxels still receive projections from the wide detector. To perform reconstruction on these voxels with limited projections, Flohr's research work completes the missing projections from the narrow detector by interpolating projections from the wide detector. Then, the final reconstruction is the weighted average of the two independent FBP reconstructions performed on the completed projections from the two detectors~\cite{Flohr08}. Such an approach not only has the same issues from the single source implementation as discussed in the previous paragraph, but can also lead to more image blurriness and artifacts from missing data interpolation and weighted averaging on independent reconstructions.

\begin{figure}
   \begin{center}
   \includegraphics[width=17cm]{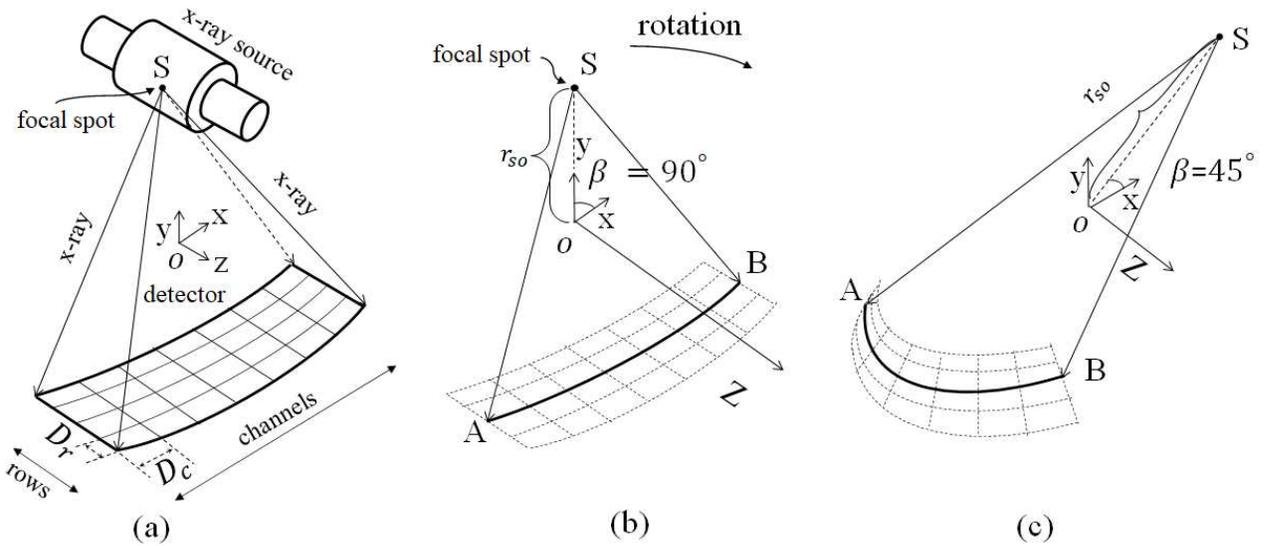}
 \caption{Single source CT scanner geometry and setup. (a) and (b) single source CT scanner at $ \ang{90}$ view angle, where S, O and $\beta$ are the focal spot, isocenter and view angle. (c) CT scanner rotated clock-wise to $\ang{45}$ view angle.   
   \label{fig:FFS1} 
    }  
    \end{center}
\end{figure}

To address the above mentioned issues, this paper proposes the JENG algorithm, which is the first physics-based statistical iterative reconstruction solution for DS-FFS CT and reconstructs from the scanner native geometry without data rebinning, interpolation or completion. Thereby, the images reconstructed by JENG have a higher spatial resolution and fewer image artifacts than the FBP methods. In addition, to avoid potential image artifacts from weighted averaging on two independent reconstructions as in the FBP methods~\cite{Flohr08,Flohr09}, the JENG algorithm uses Consensus Equilibrium to compute a single reconstruction that simultaneously fits the projections from all X-ray sources and focal spots.

\section{Materials and Methods}
\label{sec:material_methods}
\subsection{CT Setup and Math Formulation}
\label{subsec:math-formulation}

\begin{figure}[t]
   \begin{center}
    \includegraphics[width=12cm]{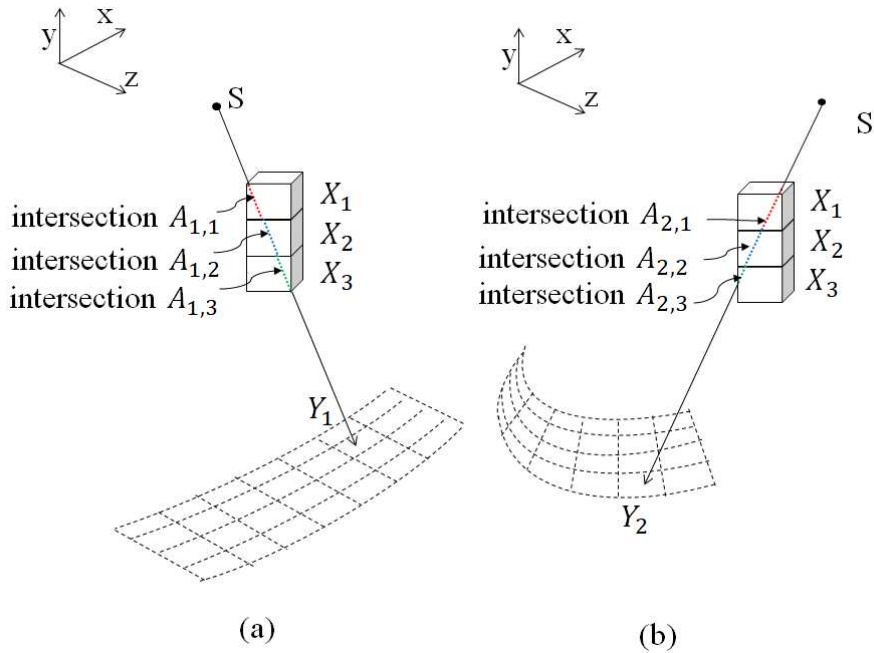}
   \caption{X-rays intersection with voxels at different view angles. (a) At $\ang{90}$ view angle, an X-ray intersects with voxels $X_1$, $X_2$ and $X_3$, and we denote the lengths of intersection between the X-ray and the voxels as $A_{1,1}$, $A_{1,2}$, and $A_{1,3}$, respectively. (b) At $\ang{45}$ view angle, a different X-ray intersects with voxels $X_1$, $X_2$ and $X_3$ with intersection lengths $A_{2,1}$, $A_{2,2}$ and $A_{2,3}$. Note that the length of intersection is unique for each voxel, each X-ray and each view angle.
   \label{fig:AMatrix} 
    }  
    \end{center}
\end{figure}

Fig.~\ref{fig:FFS1}(a) shows a CT scanner with a single X-ray source, also known as an X-ray tube, on one end of a rotating gantry, and an X-ray detector array opposite the source on the gantry. Each horizontal detector sensor unit is a channel and each vertical detector sensor unit is a row. We denote the length of each detector channel as $D_c$ and the length of each detector row as $D_r$. In addition, the total number of detector channels is $M_c$ and the total number of detector rows is $M_r$. In the example of Fig.~\ref{fig:FFS1}(a), the number of detector channels, $M_c$, is 7 and the number of detector rows, $M_r$, is 4. The center of the gantry rotation, known as the isocenter, is denoted as point O in Fig.~\ref{fig:FFS1}(a). For ease of understanding, we use a coordinate system with axis x pointing along the detector channel direction, axis y pointing upright (together x-y form the trans-axial plane), and axis z pointing along the rotation axis (axial plane). The center of the patient body to be scanned is placed near the isocenter O and the cranio-caudal direction is along the z axis. X-rays emit from a point in the X-ray source, also known as a focal spot and is denoted as S in Fig.~\ref{fig:FFS1}(a), penetrate through the patient body and project onto the X-ray detector array. Note that in this paper we symbolize the distance between focal spot, S, and isocenter, O, as $r_{so}$ and is shown in Fig.~\ref{fig:FFS1}(b). In addition, isocenter, O, is on the same plane with sector SAB, where points A and B are the two end points of the detector center row, shown as a bold arc in Fig.~\ref{fig:FFS1}(b), and sector SAB is symmetric along line SO. We also define the view angle, $\beta$, as the angle between line SO and x axis. In the example of Figs.~\ref{fig:FFS1}(a) and (b), line SO is along the y axis and the view angle $\beta$ is $\ang{90} $. When the CT scanner rotates clock-wise by $\ang{45}$ in Fig.~\ref{fig:FFS1}(c), the view angle is $\ang{45}$ in this case. 

To understand how we formulate the computations for the JENG algorithm, we use a reconstruction with three voxels, $X_1$, $X_2$ and $X_3$ as an example. Fig.~\ref{fig:AMatrix}(a) shows an X-ray intersecting with three voxels, and a detector sensor unit receives projection $Y_1$ for the X-ray at $\ang{90}$ view angle. We denote the lengths of intersection for the three voxels at the current view angle as $A_{1,1}$, $A_{1,2}$ and $A_{1,3}$ and we use different colors for each voxel's intersection length. Fig.~\ref{fig:AMatrix}(b) shows another X-ray intersecting with three voxels at $\ang{45}$ view angle with intersection lengths $A_{2,1}$, $A_{2,2}$ and $A_{2,3}$, and a different detector sensor unit takes a projection $Y_2$. Since projections $Y_1$ and $Y_2$ are the integral of radiodensity along the path of X-rays, we can express projections $Y_1$ and $Y_2$ as:
$
Y_1 = A_{1,1}X_1 + A_{1,2}X_2 + A_{1,3}X_3 +E_1\ , \quad\text{and}\quad Y_2 = A_{2,1}X_1 + A_{2,2}X_2 +A_{2,3}X_3 + E_2\ ,
$
where $X_1$, $X_2$ and $X_3$ are the radiodensity for each of the three voxels. $E_1$ and $E_2$ are the measurement errors, such as electronic and photon quantum noise, and represent the difference between measured projections, $Y_1$ and $Y_2$, and the error-free perfect projections. If we generalize the above equations for all voxels and projections, then we have:

\begin{equation}
Y = AX+E \ ,
\label{eqn:sinogram}
\end{equation}

In the above equation, $Y$ is a sinogram vector of size $M$ that includes projections from all view angles and $M$ equals $M_v \times M_c \times M_r$, where $M_v$ is the total number of view angles for the scan. $M_c$ and $M_r$ are the number of detector channels and rows as defined before.
$A$ is an $M \times N$ system matrix that models the geometry of CT, where $N$ is the size of a reconstruction. Each entry of $A$, denoted as $A_{i,j}$, represents the length of intersection between $j^{th}$ voxel and the X-ray for the $i^{th}$ sinogram entry. In addition, $A_{i,j}$ is unique for each voxel, detector sensor unit, and view angle. $X$ is a reconstruction vector of size $N$ and each element of $X$ is the radiodensity for a voxel. $E$ is a measurement error vector of size $M$, and represents the difference between $Y$ and $Y$'s error-free value. Unfortunately, we cannot directly compute reconstruction $X$ from Eqn.~(\ref{eqn:sinogram}) as measurement error $E$ is unknown and cannot be measured. In addition, inverting system matrix $A$ is impractical because an inversion takes huge amount of computations and requires terabytes of memory. To address the above challenges, the JENG algorithm computes reconstruction $X$ as the solution to the following maximum a posteriori optimization problem: 

\begin{equation}
X \gets \argmin_X \left\{ \frac{1}{2}(Y-AX)^T D (Y-AX) + R(X) \right\}  \ ,
\label{eqn:mbir}
\end{equation}
where $D$ is an $M \times M$ diagonal weight matrix and represents the inverse of the sinogram noise. $\frac{1}{2}(Y-AX)^T D (Y-AX)$ is a forward model that fits reconstruction $X$ with sinogram $Y$. If reconstruction $X$ has an anomaly, such as metal, the sinogram noise will be large and the forward model will be penalized with a small weight matrix $D$. Therefore, reconstruction $X$ has a weak fitting with beam hardened and noisy sinogram $Y$ and has less image noise or artifacts. Vice versa if the sinogram noise is small, the weight matrix $D$ is large and reconstruction $X$ has a strong fitting with noiseless sinogram $Y$. $R(X)$ in Eqn.~(\ref{eqn:mbir}) is a prior model for maintaining a good image spatial property and denoising. In this  paper $R(X)$ is a convex Generalized Markov Random Field, which denoises and penalizes each voxel based on the difference between the voxel and its neighboring voxels. A large difference leads to a strong denoising and penalization, while a small difference leads to a weak denoising and penalization. From the machine learning perspective, the forward model can be understood as the minimum mean square error of a weighted linear regression model, and $R(X)$ is a regularizer that prevents data overfitting. 

\subsection{Flying Focal Spot Geometry Modeling}
\label{subsec:FFS}
\begin{figure}[ht]
   \begin{center}
   \includegraphics[width=16.5cm]{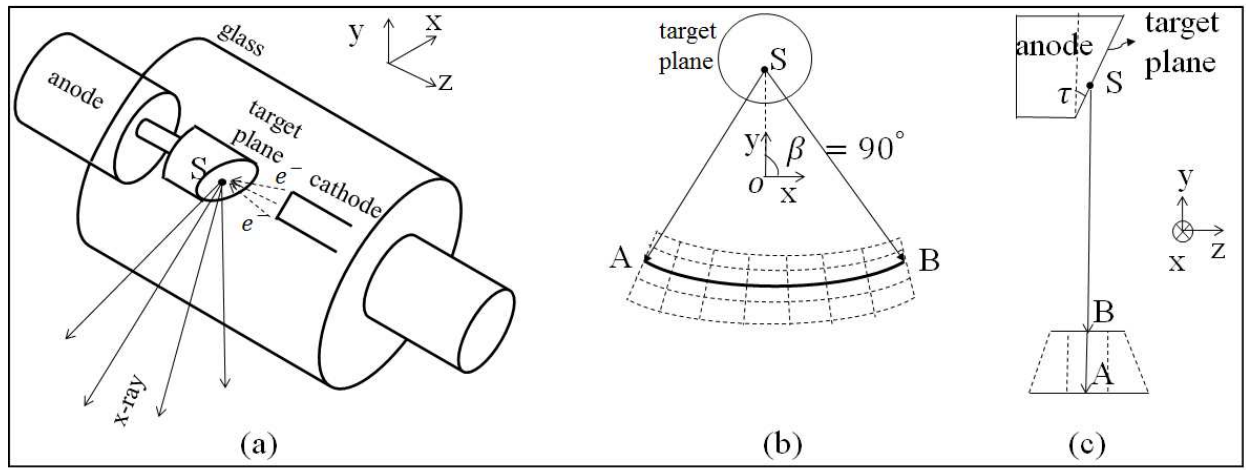}

   \caption{An X-ray source without flying focal spot. (a) Demonstrates the interior design of an X-ray source. Note that the anode target plane is tilted along the z axis. (b) The transverse view (axial plane perspective) for the focal spot and the detector at $\ang{90}$ view angle. (c) The side view (sagittal plane perspective) for the focal spot and the detector.
   \label{fig:FFS2} 
    }  
 \end{center}
\end{figure}

\begin{figure}[ht]
   \begin{center}
    \includegraphics[width=15cm]{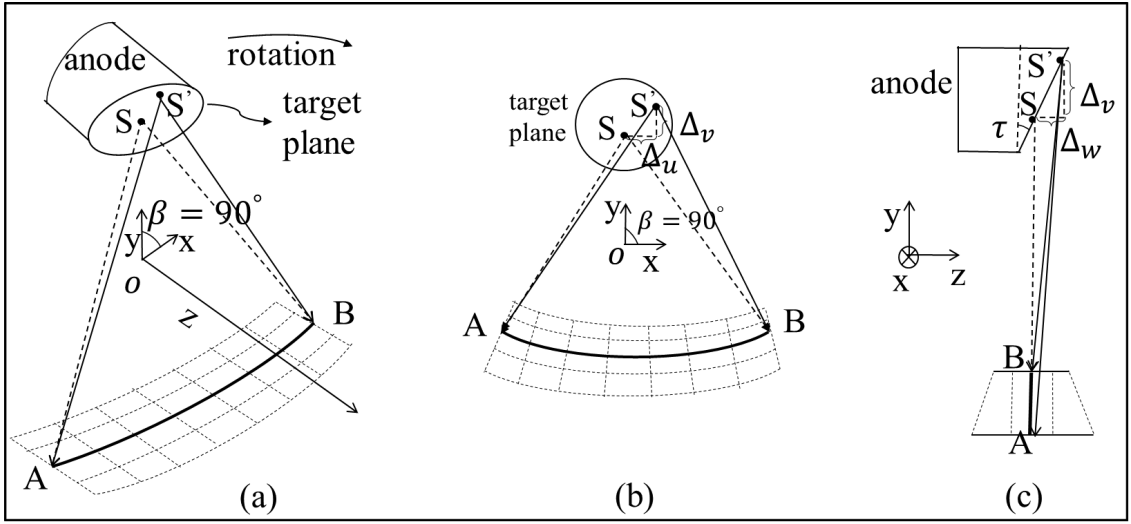}
   \caption{An X-ray source with flying focal spots. (a) Shows a single source CT setup with two focal spots, S and S\textsuperscript{'}. (b) Shows the axial plane transverse view of the focal spots at $\ang{90}$ view angle. (c) The sagittal plane side view of the focal spots. Note that S and S\textsuperscript{'} are on the same tilted target plane.
   \label{fig:FFS3} 
    }  
    \end{center}
\end{figure}

To compute system matrix $A$ in Eqn.~(\ref{eqn:mbir}), we start by examining the structure inside the X-ray source through Fig.~\ref{fig:FFS2}(a).
The X-ray source consists of an encapsulating glass envelope, a rotating anode, a cathode and a tilted target plane. The target plane is also connected with the anode and is tilted from the X-Y plane by an anode tilt angle $\tau$. 
From the cathode, the accelerated electrons, denoted as $e^{-}$ in Fig.~\ref{fig:FFS2}(a), hit the focal spot S on the anode target plane and the X-rays are then produced at the focal spot. Figs.~\ref{fig:FFS2}(b) and (c) show the X-ray source structure at $\ang{90}$ view angle from transverse view (axial plane) perspective and side view (sagittal plane) perspective and we can observe that focal spot S lies on the tilted target plane. Fig.~\ref{fig:FFS2}(c) also shows the anode angle $\tau$ as the angle between y axis and the tilted target plane. Furthermore to the above introduction, the precise coordinate location $(x_s,y_s,z_s)$ for focal spot S at any view angle $\beta$ can be computed as:

\begin{equation}
\centering
x_s =  r_{so}\cos (\beta) \ , \quad\text{and}\quad
y_s =  r_{so}\sin (\beta)  \ , \quad\text{and}\quad 
z_s =  \frac{H_{r}(\beta - \beta_0)}{2\pi} \ ,  \label{eqn:ffs_coordinate} 
\end{equation}
where the cosine and sine trigonometry relationship for $x_s$ and $y_s$ can be observed in Fig.~\ref{fig:FFS1}(c). $H_r$ is how far the X-ray source moves along the z axis in a $\ang{360}$ rotation, $\beta$ is the view angle in radian, and $\beta_0$ is the CT scanner view angle when $z_s=0$.

When the scanner is equipped with flying focal spot, the cathode quickly wobbles and electrons hit the target plane at multiple different locations, creating multiple X-ray focal spots at each view angle. Each focal spot produces a set of projections and different focal spots produce different but interleaved projections. By producing conjugate sets of projections through flying focal spots, the total number of projections, $M$, increases proportionally with the number of focal spots but the CT scan time does not significantly lengthen. Fig.~\ref{fig:FFS3}(a) shows an example single source CT scanner with two focal spots, the default focal spot S and the deflected focal spot S\textsuperscript{'}, and Figs.~\ref{fig:FFS3}(b) and (c) show S and S\textsuperscript{'} in the axial and sagittal plane perspectives at $\ang{90}$ view angle. We can observe from Figs.~\ref{fig:FFS3}(a) and (b) that projections from S and S\textsuperscript{'} overlap but do not contain each other.

To compute the coordinate location for flying focal spot, we denote the displacement vector between S and S\textsuperscript{'} at $\ang{90}$ view angle as $( \Delta_u , \Delta_v , \Delta_w)$ in Figs.~\ref{fig:FFS3}(b) and (c). Since focal spots S and S\textsuperscript{'} are on the same target plane, $\Delta_v$ and $\Delta_w$ have the following trigonometry relationship: $\Delta_w = \tan(\tau)  \Delta_v$, where $\tau$ is the anode tilt angle. The deflected focal spot, S\textsuperscript{'}, at any view angle $\beta$ can then be given below with coordinate $(x_s^{'} , y_s^{'} , z_s^{'})$:

\begin{align}
\left[
\begin{array}{c}
x_s^{'} \\
y_s^{'}  \\
z_s^{'}
\end{array}
\right]  =  \left[
\begin{array}{c}
 r_{so}\cos (\beta) \\
r_{so}\sin (\beta)  \\
\frac{H_{r}(\beta - \beta_0)}{2\pi}
\end{array}
\right]\  +  R_\beta \left[\begin{array}{c}
\Delta_u\\
\Delta_v  \\
\tan(\tau)  \Delta_v
\end{array}
\right] \ , \quad\text{and}\quad R_\beta =
\left[
\begin{array}{ccc}
\sin(\beta) & \cos(\beta) & 0 \\
-\cos(\beta) & \sin(\beta) & 0 \\
0 & 0 & 1 
\end{array} 
\right] \ ,
\label{eqn:deflected_focal_equation}
\end{align}
where the vector before the addition operator is the coordinate location for the default focal spot from Eqn.~(\ref{eqn:ffs_coordinate}). $R_\beta$ is a rotation matrix that rotates the displacement between S\textsuperscript{'} and S at $\ang{90}$ view angle to any view angle, $\beta$, assuming that a helical CT scanner rotates in the x-y plane and translates in the z direction.
After plugging in the $R_\beta$ expression,  Eqn.~(\ref{eqn:deflected_focal_equation}) can be reorganized as:

\begin{align}
\left[
\begin{array}{c}
x_s^{'} \\
y_s^{'}  \\
z_s^{'}
\end{array}
\right]  =  \left[
\begin{array}{c}
 r_{so}\cos(\beta) + \sin(\beta)\Delta_u +\cos(\beta)\Delta_v\\
r_{so}\sin(\beta) - \cos(\beta)\Delta_u + \sin(\beta)\Delta_v  \\
\frac{H_{r}(\beta - \beta_0)}{2\pi} + \tan(\tau)  \Delta_v
\end{array}
\right]\  \ ,
\label{eqn:deflected_focal_equation2}
\end{align}
Note that Eqn.~(\ref{eqn:deflected_focal_equation2}) is a general form for the coordinate location of any focal spot, with or without deflection. If $\Delta_u$ and $\Delta_v$ are both zeros, then Eqn.~(\ref{eqn:deflected_focal_equation2}) is the default focal spot coordinate without deflection and is the same as Eqn.~(\ref{eqn:ffs_coordinate}). Readers should also know that Eqn.~(\ref{eqn:deflected_focal_equation2}) assumes that a focal spot is a sizeless point without actual physical shape. Given that CT scans often use a small focal spot with size less than 1mm for optimal diagnostic values, the focal spot size approximation in Eqn.~(\ref{eqn:deflected_focal_equation2}) has minimal or no impact on spatial resolution. In the unusual cases with a focal spot size larger than 1mm, the sizeless point assumption might lead to sub-optimal spatial resolution.   

Knowing the coordinate for the focal spots alone, however, is not sufficient to compute system matrix entry, $A_{i,j}$. We also need to know the geometry information for voxel $X_j$ and we introduce two other parameters in this paper, $\theta$ and $\phi$.
$\theta$ is voxel $X_j$'s ray angle in the x-y plane parallel to the x axis, and Fig.~\ref{fig:FFS4}(a) depicts $\theta$ in the x-y plane. Fig.~\ref{fig:FFS4}(a) also denotes point $C$ as the location where the X-ray hits the detector in the x-y plane. $\phi$ is the voxel's ray angle in the y-z plane and parallel to the line connecting S\textsuperscript{'} and $C$, and is shown in Fig.~\ref{fig:FFS4}(b). With the above definition, $\theta$ and $\phi$ can then be computed from the below trigonometry equations:

\begin{align}
   \centering
\theta = \arctantwo \left( \frac{y_s^{'} -y_j}{x_s^{'} -x_j} \right) \ ,  \quad
\phi =  \arctantwo \left( \frac{z_s^{'} -z_j}{\sqrt{(x_s^{'}-x_j)^2 + (y_s^{'}-y_j)^2}} \right) \ , \label{eqn:theta_phi}  
\end{align}
where $\arctantwo$ operator returns the arctangent value in the range of $\left[ -\pi, \pi \right]$, $(x_s^{'},y_s^{'},z_s^{'})$ is the focal spot coordinate location from Eqn.~(\ref{eqn:deflected_focal_equation2}), and $(x_j,y_j,z_j)$ is voxel $X_j$'s coordinate location.
To ensure that the length of intersection, $A_{i,j}$, is never negative, we introduce two more parameters, $\tilde{\theta}$ and $\tilde{\phi}$, which are $\ang{45}$ rotations of $\theta$ and $\phi$, and we clip their values to$\left[-\frac{\pi}{4}, \frac{\pi}{4} \right]$.  $\tilde{\theta}$ and $\tilde{\phi}$ are defined below as:
\begin{figure}[t]
   \begin{center}
    \includegraphics[width=16.5cm]{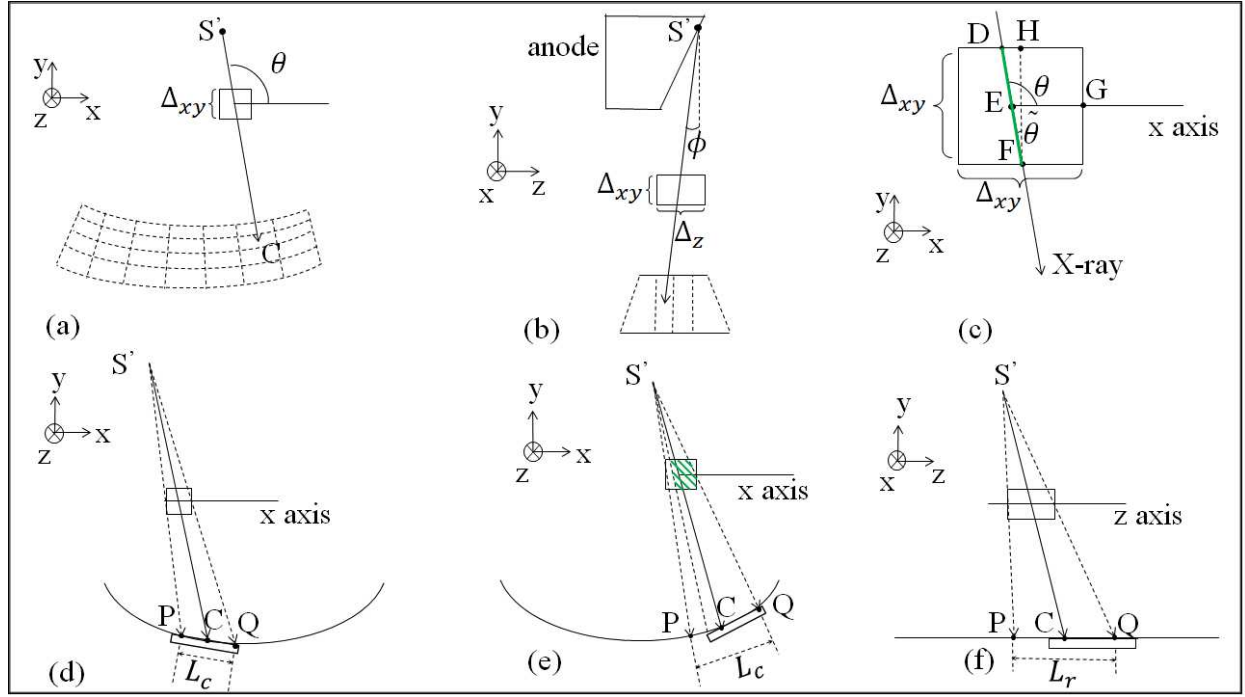}
   \caption{(a) and (b) show x-y plane ray angle $\theta$ and y-z plane ray angle $\phi$. (c) Green line segment DF is the length of intersection between a voxel and an X-ray that goes through the voxel center. (d) Shows a voxel whose projection onto the detector (x-y plane) is completely within a channel. (e) The voxel's projection partially overlaps with the detector channel. (f) The voxel's projection onto the detector (y-z plane) partially overlaps with a row.
   \label{fig:FFS4} 
    }  
    \end{center}
\end{figure}

\begin{align}
   \centering
\tilde{\theta} = \left[\left( \theta +\frac{\pi}{4} \right)\bmod \frac{\pi}{2} \right] -\frac{\pi}{4} \ , \quad
\tilde{\phi} = \left[\left( \phi +\frac{\pi}{4} \right)\bmod \frac{\pi}{2} \right] -\frac{\pi}{4}  \ , \label{eqn:theta_phi_new}  
\end{align}
With  $\tilde{\theta}$ and $\tilde{\phi}$, the length of intersection, $A_{i,j}$, can then be computed as:

\begin{equation}
A_{i,j} =  \frac{\Delta_{xy}}{\cos{\tilde{\theta}}\cos{\tilde{\phi}}}
\label{eqn:A_ij}
\end{equation}

In the above equation, we denote a voxel's x-y plane side length as $\Delta_{xy}$ and z direction side length as $\Delta_z$ in Figs.~\ref{fig:FFS4}(a) and (b). To explain how Eqn.~(\ref{eqn:A_ij}) is derived, Fig.~\ref{fig:FFS4}(c) shows an X-ray intersecting with voxel $X_j$ and the intersection length, line segment DF, is colored in green. Point E in Fig.~\ref{fig:FFS4}(c) is the center of the voxel with points D, E, F along the same line. The length of line segment FH equals $\Delta_{xy}$ and FH is parallel to y axis with point H on the edge of the voxel. $\sphericalangle$DEG is ray angle $\theta$ and $\sphericalangle$DFH is $\tilde{\theta}$, such that $\theta$ and $\tilde{\theta}$ are $\ang{90}$ apart in this example. Given that both the length of line segment FH and $\sphericalangle$DFH are known, the length of X-ray's intersection in the x-y plane, segment DF, can be computed as $\frac{\Delta_{xy}}{\cos{\tilde{\theta}}}$. To project line segment DF from the x-y plane to a three-dimensional space, Eqn.~(\ref{eqn:A_ij}) divides DF segment length with $\cos{\tilde{\phi}}$.

Eqn.~(\ref{eqn:A_ij}), however, assumes that a voxel's projection is entirely taken by a single detector sensor unit. When a voxel's projection is taken by multiple detector sensor units, the measurement from any single detector sensor unit, $Y_i$, can no longer account for the entire voxel's projection. Instead, $Y_i$ is the projection for the portion of the voxel that is in the way of the X-rays from the focal spot to the detector sensor unit  that receives $Y_i$. In the example of Fig.~\ref{fig:FFS4}(d), the voxel's projection in the x-y plane, shown as line segment PQ in the figure with length $L_c$, is entirely within a channel. In this example, the projection taken by the detector channel accounts for the entire voxel and the computations for Eqn.~(\ref{eqn:A_ij}) is accurate. In another example in Fig.~\ref{fig:FFS4}(e), the voxel is partially in the way of the X-rays and its projection in the x-y plane, namely line segment PQ, does not fully overlap with the detector channel. In such an example, the projection taken by the channel only accounts for the portion of the voxel shaded in green and Eqn.~(\ref{eqn:A_ij}) no longer holds. Similarly, Fig.~\ref{fig:FFS4}(f) gives an example voxel whose projection in the y-z plane, line segment PQ with length $L_r$, does not fully overlap with the detector row.

\begin{figure}[t]
   \begin{center}
   \includegraphics[width=16.5cm]{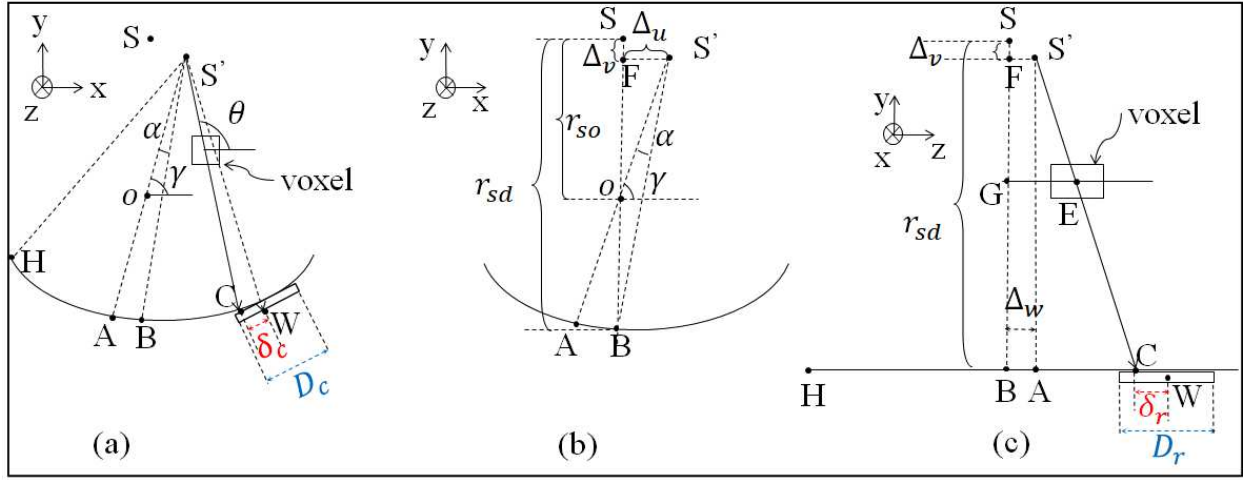}
   \caption{(a) and (b) demonstrate how $\delta_c$ is computed, with $\delta_c$ colored as a red segment between points C and W. (c) Shows how $\delta_r$ is computed. Similar to (a), $\delta_r$ is red segment CW.\label{fig:FFS5}}
 \end{center}
\end{figure}

Therefore, $A_{i,j}$ in Eqn.~(\ref{eqn:A_ij}) must be modified so that $A_{i,j}$ not only reflects the length of intersection between X-rays and voxels, but also reflects the overlap between the voxel's projection and the corresponding detector sensor unit.  To do so, $A_{i,j}$ is multiplied by a normalization term and is computed in the following way:  
\begin{equation}
 A_{ij} =  \frac{\Delta_{xy}}{\cos{\tilde{\theta}}\cos{\tilde{\phi}}} \left[  V( \delta_c) * W( \delta_c ) \right] \times  \left[V( \delta_r) * W( \delta_r ) \right] \ ,
\label{eqn:A_ij_normalized}
\end{equation}
where $\delta_c$ is the x-y plane displacement between the center of voxel $X_j$'s projection and the center of the channel that receives projection $Y_i$. $\delta_r$ is the y-z plane displacement between the center of the voxel's projection and the center of the detector row. Figs.~\ref{fig:FFS5}(a), (b) and (c) depict $\delta_c$ and $\delta_r$ as red segments CW, where point C is the center of the voxel's projection and point W is the center of the corresponding detector sensor unit.

In addition to the above notations, $V(\cdot)$ in Eqn.~(\ref{eqn:A_ij_normalized}) is a voxel density function and $W(\cdot)$ is a detector sensitivity function. $*$ is a convolution operation and $\times$ is a multiplication operation. $\frac{\Delta_{xy}}{\cos{\tilde{\theta}}\cos{\tilde{\phi}}}$ computes the length of intersection between the X-ray and voxel $X_j$, as explained in Eqn.~(\ref{eqn:A_ij}), whereas $\left[  V( \delta_c) * W( \delta_c ) \right] \times  \left[V( \delta_r) * W( \delta_r ) \right]$ is a normalization term that accounts for the overlap between voxel $X_j$'s projection and the detector sensor unit that receives the projection. $  V( \delta_c) * W( \delta_c )$ accounts for the x-y plane overlap and $V( \delta_r) * W( \delta_r ) $ accounts for the y-z plane overlap. When a voxel's projection completely overlaps with a detector sensor unit, such as in the example of Fig.~\ref{fig:FFS4}(d), the normalization term is 1. Otherwise, the normalization term is between 0 and 1.
To define $V(\delta_c)$, $V(\delta_r)$, $W(\delta_c)$, and $W(\delta_r)$, we assume that each voxel has a uniform radiodensity everywhere in the voxel and each detector sensor unit has a uniform sensitivity. For simplicity, we borrow the voxel density and detector sensitivity functions from citation~\cite{Thibault07}, and define $V(
\cdot)$ and $W(\cdot)$ as the following rectangular functions:

\begin{align}
 V(\delta_c) &= \textbf{rect}\left( \frac{\delta_c}{L_c}\right) \ , \quad\text{and}\quad  V(\delta_r) = \textbf{rect}\left( \frac{\delta_r}{L_r}\right) \,  \label{eqn:V_Function} \\ 
  W(\delta_c) &= \frac{1}{D_c}\textbf{rect}\left( \frac{\delta_c}{L_c}\right) \ , \quad\text{and}\quad  W(\delta_r) = \frac{1}{D_r}\textbf{rect}\left( \frac{\delta_r}{L_r}\right) \label{eqn:S_Function}
\end{align}
where $L_c$ and $L_r$, as defined before, are the lengths of the voxel's projection in the x-y plane and y-z plane, and are represented as line segment PQ in Figs.~\ref{fig:FFS4}(d), (e) and (f). $D_c$ and $D_r$ are the lengths for a detector channel and row, respectively, and are colored as blue line segments in Figs.~\ref{fig:FFS5}(a) and (c). After plugging Eqns.~(\ref{eqn:V_Function}) and~(\ref{eqn:S_Function}) into Eqn.~(\ref{eqn:A_ij_normalized}), we eliminate the convolution operations and rewrite Eqn.~(\ref{eqn:A_ij_normalized}) in a closed-form expression: \useshortskip

 \begin{equation}
 A_{ij} = \frac{\Delta_{xy}}{D_c D_r\cos{\tilde{\theta}}\cos{\tilde{\phi}}} \textbf{clip} \left[ 0 , \frac{D_c + L_c}{2} - \lvert \delta_c\rvert, \min \left( L_c, D_c\right)\right]  \times 
  \textbf{clip} \left[ 0 , \frac{D_r + L_r}{2} - \lvert \delta_r\rvert, \min \left( L_r, D_r\right)\right] \ ,
\label{eqn:A_ij_new}
\end{equation}
with function \textbf{clip} defined as: \textbf{clip}$[a, b, c] =\min\left(\max\left(a,b\right),c\right)$.
To compute $A_{i,j}$ correctly, $\delta_c$ and $\delta_r$ in the above equation must compensate for the flying focal spot deflection. The value for $\delta_c$ is clipped to $[-(r_{sd}+\Delta_v)\pi, (r_{sd}+\Delta_v)\pi )$ and can be computed as:

 \begin{equation}
\delta_c  \approx \left[\left(\theta - \gamma  -\alpha + \frac{(M_c -1)D_c}{2(r_{sd}+\Delta_v)} -\frac{i_c D_c}{r_{sd}+\Delta_v}+\pi \right)\bmod 2\pi -\pi \right] \left(r_{sd}+\Delta_v \right) \ ,
\label{eqn:delta_c}
\end{equation}
where $\gamma$ and $\alpha$ are defined as:
\begin{align}
\gamma &= \beta - \arctantwo \left( \frac{\Delta_u}{r_{so}+\Delta_v} \right)
 \\
\alpha &= \arctantwo \left(\frac{r_{sd}+\Delta_v}{\Delta_u} \right) - \arctantwo \left(\frac{r_{so}+\Delta_v}{\Delta_u}\right) 
\label{eqn:alpha}
\end{align}

In the above equations, $\theta$ is the X-ray ray angle in the x-y plane as defined before; $\gamma$ is the angle between S\textsuperscript{'}O and x axis, and both $\theta$ and $\gamma$ are pointed out in Figs.~\ref{fig:FFS5}(a) and (b). $\alpha$ is the detector channel offset, defined as the angle between the X-ray hitting the detector center in the x-y plane and the X-ray passing through the isocenter, O. In the example of Figs.~\ref{fig:FFS5}(a) and (b), $\alpha$ is $\sphericalangle$AS\textsuperscript{'}B. Point A is where the X-ray through isocenter O hits the detector array in the x-y plane, and point B is the detector array center in the x-y plane. $M_c$ is the total number of detector channels as defined in Sec.~\ref{subsec:math-formulation}. $r_{so}$ is the distance from the default focal spot, S, to the isocenter, O. $r_{sd}$ is the vertical distance from S to the detector array, and both $r_{so}$ and $r_{sd}$ are indicated in Fig.~\ref{fig:FFS5}(b). $\Delta_u$ and $\Delta_v$ are the focal spot location displacement, along x axis and y axis respectively, from  S to the deflected focal spot, S\textsuperscript{'}, and are both indicated in Fig.~\ref{fig:FFS5}(b). $i_c$ is the index for the detector channel that receives voxel $X_j$'s projection, and we assume that the index for the leftmost detector channel is zero. In the example of Figs.~\ref{fig:FFS5}(a) and (b), we use point H to represent the center for the leftmost detector channel
and point W for the center of the ${\it i_c}^{th}$ channel. 

In Eqn.~(\ref{eqn:delta_c}), operation $\theta - \gamma$ computes angle $\sphericalangle$AS\textsuperscript{'}C in Fig.~\ref{fig:FFS5}(a). By subtracting $\alpha$, which is $\sphericalangle$AS\textsuperscript{'}B, the  first three terms, $\theta - \gamma - \alpha$, compute $\sphericalangle$BS\textsuperscript{'}C. The fourth term, $\frac{(M_c -1)D_c}{2(r_{sd}+\Delta_v)}$, is the approximated angle HS\textsuperscript{'}B. 
Therefore, The result of computing the first four terms in Eqn.~(\ref{eqn:delta_c}) is $\sphericalangle$ HS\textsuperscript{'}C in the example of Fig.~\ref{fig:FFS5}(a). The fifth term is the approximated angle $\sphericalangle$HS\textsuperscript{'}W. Together, the computations for the first five terms in Eqn.~(\ref{eqn:delta_c}) get $\sphericalangle$CS\textsuperscript{'}W, which is the angular measure for $\delta_c$, by subtracting $\sphericalangle$HS\textsuperscript{'}C from $\sphericalangle$HS\textsuperscript{'}W. Then, Eqn.~(\ref{eqn:delta_c}) clips the angle to the range of $[-\pi ,\pi)$ and converts from angular measure to arc length at the end of the equation. Similarly, we can compute $\delta_r$ as:
 \begin{equation}
\delta_r = \frac{r_{sd}+\Delta_v}{\sqrt{(x_s^{'}-x_j)^2 +(y_s^{'}-y_j)^2}}(z_j-z_s^{'})+\frac{M_r-1}{2}D_r + \Delta_w - i_r D_r \ ,
\label{eqn:delta_r}
\end{equation} 
where $(x_s^{'}, y_s^{'}, z_s^{'})$ is the coordinate location for a deflected focal spot S\textsuperscript{'} and $(x_j,y_j,z_j)$ is the coordinate location for voxel $X_j$. $M_r$ is the number of detector rows. $\Delta_w$ is the focal spot displacement along z axis from S to  S\textsuperscript{'} and is indicated in Fig.~\ref{fig:FFS5}(c). $i_r$ is the index for the detector row that receives a projection for voxel $X_j$ and we assume that the leftmost row has index 0. In the example of Fig.~\ref{fig:FFS5}(c), we use point H to indicate the center for the leftmost detector row. The first term in Eqn.~(\ref{eqn:delta_r}) computes length AC, where point A is on the detector rows and line S\textsuperscript{'}A is parallel to y axis. The second term computes length HB, where point B is the center of the detector array in the y-z plane. The third term, $\Delta_w$, equals to the length of AB, and the first three terms together compute length HC. The fourth term, $i_rD_r$, computes length HW. In the end, Eqn.~(\ref{eqn:delta_r}) computes $\delta_r$ by subtracting length HC from length HW.

\subsection{Dual Source CT Modeling and Computations}
\label{subsec:dual-source}
\begin{figure}[ht]
   \begin{center}
   \includegraphics[width=14cm]{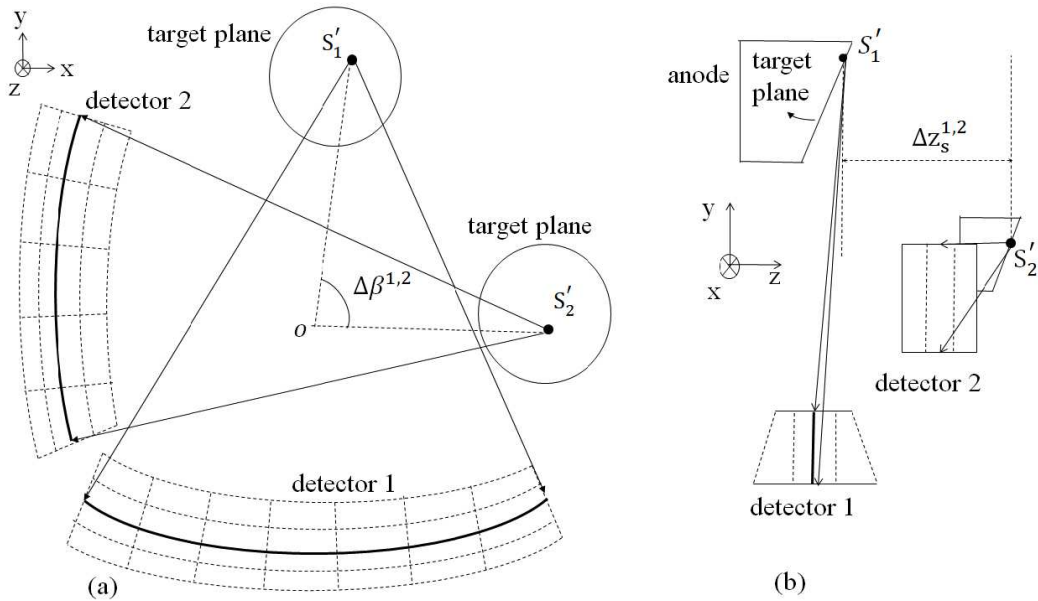}
 \caption{ Dual source CT design and geometry. (a) Shows detector 1 of a dual source CT covering the full field of view, and detector 2 covering a smaller and central field of view. In addition, the two detectors are offset by a constant rotation angle, $\Delta \beta^{1,2}$, in the x-y plane. (b) The two detectors are offset by a displacement $\Delta z^{1,2}$ in the y-z plane.
   \label{fig:DS-CT} 
    }  
    \end{center}
\end{figure}

For a dual source CT design, two sources and two detectors at the same X-ray energy level are mounted on a rotating gantry, with each pair of source and detector acquiring conjugate but different projections. In addition, different pairs have different geometry parameters with different detector sizes, view angles, and X-ray source movement. 
Fig.~\ref{fig:DS-CT}(a) shows an example dual source CT. The first X-ray source is S\SPSB{'}{1} and its corresponding detector has 7 channels and a large field-of-view. The second X-ray source is S\SPSB{'}{2} and its corresponding detector has a smaller size with 5 channels and a smaller field-of-view, given that the rotating gantry has limited space to fit two full size detectors. The focal spot locations between S\SPSB{'}{1} and S\SPSB{'}{2} are offset by an angular displacement of $\Delta \beta^{1,2}$ in the x-y plane and a translation displacement of $\Delta z_s^{1,2}$ in the z direction. Both $\Delta \beta^{1,2}$ and $\Delta z_s^{1,2}$ are shown in Fig.~\ref{fig:DS-CT} and $\Delta \beta^{1,2}$ is represented by $\sphericalangle$ S\SPSB{'}{1}OS\SPSB{'}{2} in Fig.~\ref{fig:DS-CT}(a). In practice, $\Delta \beta^{1,2}$ is often chosen to be 90\si{\degree} for an efficient mechanical assembly of the detector sensor units.

\begin{figure}[ht]
   \begin{center}
   \includegraphics[width=16.5cm]{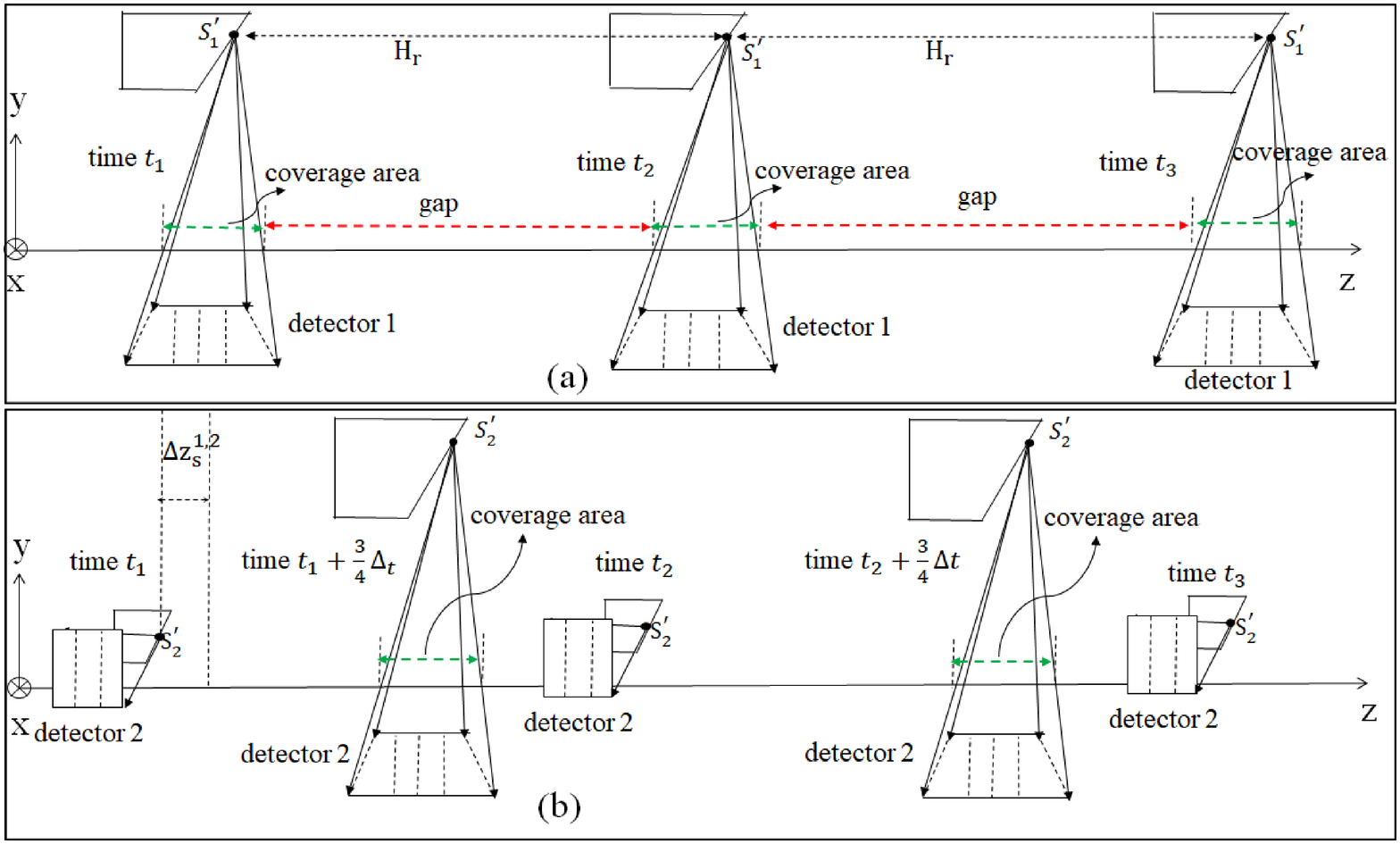}
 \caption{Demonstrates the dual sources' coordinated movement and shows their X-ray coverage gaps and overlaps. (a) Dual Source CT detector 1's movement in the z direction across time. Time $t_1$, $t_2$ and $t_3$ are three consecutive rotations. Note that the gaps between consecutive rotations are much larger than the X-ray coverage areas. (b) Detector 2's z axis movement at time $t_1$, $t_2$ and $t_3$. Note that detector 2's X-ray coverage areas partially fill the coverage gaps of detector 1.}
   \label{fig:DS-gap} 
    \end{center}
\end{figure}

With the above design, the dual source CT has the unique advantage to enable high pitch scans and rapid data acquisition without significantly increasing undersampling image artifacts. To explain why, Fig.~\ref{fig:DS-gap} demonstrates the interleaved helical trajectory of a dual source CT. Assuming that the scan has a high pitch, Fig.~\ref{fig:DS-gap}(a) shows detector 1's movement along the z axis across time. Time $t_1$, $t_2$ and $t_3$ are the 90\si{\degree} view angles from three consecutive rotations when detector 1 and its source are in an upright position. In addition, it takes exactly $\Delta_t$ seconds and a distance of $H_r$ for the X-ray source to move from its location at time $t_1$ to $t_2$ or $t_2$ to $t_3$, where $H_r$ is the rotation distance and was defined before in Eqn.~(\ref{eqn:ffs_coordinate}). At 90\si{\degree} view angles, the areas of the patient body covered by X-rays are marked by green line segments. Meanwhile, the high pitch scan leads to X-ray coverage gaps between consecutive rotations, shown as red line segments in the figure, and the gaps are not covered by any X-rays at 90\si{\degree} view angles. Consequently, a reconstruction from detector 1's projections alone will lead to unacceptable image quality with significant undersampling artifacts. To reduce the artifacts, detector 2 and its source are designed to provide extra X-rays coverage that diminishes the coverage gaps of detector 1 and increases the number of  projection samplings. In Fig.~\ref{fig:DS-gap}(b), detector 2 and its source have a 90\si{\degree} angular displacement from detector 1 and lie horizontally at time $t_1$. In addition, detector 2 have a z direction translation displacement of $\Delta z_s^{1,2}$ from detector 1. From time $t_1$, it takes detector 2 and its source exactly $\frac{3}{4}\Delta_t$ seconds to rotate to an upright position. In the example of Fig.~\ref{fig:DS-gap}(b), we note that along the z axis, detector 2 at time $t_1 + \frac{3}{4}\Delta_t$ is located between the positions for detector 1 at time $t_1$ and $t_2$. Therefore, the X-ray coverage for detector 2 partially fills the gaps of detector 1 and provides the missing projections for detector 1 in areas where no X-rays is available. With the additional projections from detector 2, a dual source CT thereby minimizes the increase in high pitch undersampling artifacts.

Despite of the benefits above, we can observe from Fig.~\ref{fig:DS-gap} that the additional projections from detector 2 do not completely fill the X-rays coverage gaps of detector 1. Therefore, a high pitch dual source CT can still lead to noticeable undersampling artifacts in reconstructed images due to the projection undersampling. To make it even worse, the state-of-the-art FBP methods for dual source CT heavily utilize data interpolation and geometry approximation, leading to even more pronounced undersampling artifacts~\cite{Flohr08,Flohr09}. 
In response, this section proposes a joint estimation framework for JENG that has no data interpolation and the reconstruction simultaneously fits projections and geometry from both source-detector pairs. Consequently, JENG can better take advantage of the dual source CT design for more effective artifacts reduction than FBP. To implement the JENG algorithm, we construct the system matrices $A^1$ and $A^2$ for the first and the second source-detector pairs by following the system matrix computations in Eqn.~(\ref{eqn:A_ij_new}). Then, the following joint estimation cost function finds a consensus solution that fits projections and geometry for both source-detector pairs:
\begin{equation}
X \gets \argmin_X \left\{ \frac{1}{2} \norm{Y^1 - A^1 X}^2_{D^1} + \frac{1}{2} \norm{Y^2 - A^2 X}^2_{D^2} + R(X) \right\}  \ ,
\label{eqn:dual-source}
\end{equation}
where $Y^1$, $A^1$, and $D^1$ are the sinogram projections, system matrix and weight matrix for the first source-detector pair, and $Y^2$, $A^2$, and $D^2$ are those for the second pair. $X$ is a consensus reconstruction that fits geometries and projections from both pairs. $\norm{Y^1 - A^1 X}^2_{D^1}$ is a short-hand notation for vector norm $(Y^1-A^1X)^T D^1 (Y^1-A^1X)$. Similarly, the same notation is applied to $\norm{Y^2 - A^2 X}^2_{D^2}$.

To consider cases when each source-detector pair has multiple flying focal spots, we construct an independent system matrix for each focal spot at each source-detector pair. For example, if a dual source CT has four focal spots at each source-detector pair, then we have eight system matrices in total. In a general case with $K$ system matrices, the joint estimation cost function has the following form:

\begin{equation}
X \gets \argmin_X \left\{ \frac{1}{2} \sum_{k=1}^{K}\norm{Y^k - A^k X}^2_{D^k}  + R(X) \right\}  \ ,
\label{eqn:DS-FFS-CT}
\end{equation}
where $K$ is the total number of system matrices that equals to the number of source-detector pairs multiplied by the number of focal spots, and $k$ is the index for $k^{th}$ system matrix and forward model. $Y^k$, $D^k$ are the projections and weight matrix corresponding to the $k^{th}$ system matrix.

As all forward and prior models in the above equation are strictly convex, there exists a unique global minimum to Eqn.~(\ref{eqn:DS-FFS-CT}). A wide variety of numerical methods can be used to compute the global minimum of the above cost function, including Iterative Coordinate Descent~\cite{Sauer93,XWang16, Sabne17,XWang17-2}, Gradient Descent, and Conjugate Gradient Descent~\cite{Shewchuk94}. Unfortunately no matter which numerical method to use, all methods have trouble with storing the system matrices in memory as the system matrices have a very large memory requirement and their memory size is proportional to both the reconstruction size, $N$, and the measurements size, $M$. For a large-scale reconstruction, the system matrices can sometimes take hundreds of terabytes of memory, which is beyond what a standard computer workstation can possibly provide~\cite{XWang19}. 

To lower the memory requirement, there exists two approaches: (1) the ordered-subsets method~\cite{Erdogan,Fessler03}, and (2) the on-the-fly method~\cite{Qi05}.
The ordered-subsets method splits measurements and the system matrices into subsets and distributes
them among compute nodes. Therefore, the memory requirement for each node is the assigned subset only. Each node then computes a private reconstruction using its assigned subset and merges the private reconstructions
from all nodes into a consensus solution. The ordered subsets method, however, is an approximation method and a convergence to the global minimum to Eqn.~(\ref{eqn:DS-FFS-CT}) is not guaranteed~\cite{Erdogan}. If measurements are uniformly sampled among subsets, the ordered-subset method
converges to a value close to the global minimum. If measurements are partitioned non-uniformly, the converged solution is far from the global minimum.
In contrast, the on-the-fly method divides voxels into small groups and updating all groups in sequence. At each voxel group, the on-the-fly method computes and stores a small portion of the system matrices that are needed by the current group. Once the on-the-fly method finishes the update for the current group and moves on to the next one, the previous memory storage is emptied and a new portion of the system matrices for the next voxel group are computed and stored in memory. Despite that the on-the-fly method has little memory requirement, the on-the-fly method significantly increases the computation cost and reconstruction time because the on-the-fly method recomputes the entire system matrices in every iteration of the numerical method. 

Recently, a new numerical method, Consensus Equilibrium (CE), partitions the system matrices and measurements across nodes in \ul{\textbf{any}} order, including any non-uniform sampling~\cite{Buzzard17,XWang19,Sridhar19}. Then each node computes an individual reconstruction from its assigned subset and merges individual reconstructions into a consensus solution that is provably exactly the global minimum to  Eqn.~(\ref{eqn:DS-FFS-CT}). Therefore unlike the ordered-subsets approximation method, CE is a precise method and convergence is guaranteed for any partition.
In addition, the CE Method pre-computes and stores partitioned system matrices and measurements in the memory of each node. Therefore, the CE Method avoids problems such as repeated computations and reconstruction time increase that are prevalent for the on-the-fly method. Although the CE Method has clear advantages over both the ordered-subsets and the on-the-fly method, the CE Method has not been used for DS-FFS CT iterative reconstruction and this section discusses how the CE Method is used for such purpose for the first time. 

To understand the CE Method, we use notation $V^k$ for the individual reconstruction that fits projection subset $Y^k$ and system matrix subset $A^k$. Then the CE Method's proximal function, denoted as $F_k(X)$ and defined below, finds a balance between individual reconstruction, $V^k$, and the consensus solution, $X$:

\begin{equation}
V^k \gets \argmin_{V^k}F_k(X) = \argmin_{V^k} \left\{ \frac{1}{2} \norm{Y^k - A^k V^k}^2_{D^k}  + \frac{R(V^k)}{K} +\frac{\norm{ V^k - X}^2}{2 \sigma^2} \right\}  \ ,
\label{eqn:CE}
\end{equation}
where the first two terms for $F_k(X)$ fit the individual reconstruction, $V^k$, with the $k^{th}$ forward model and the prior model. The third term, $\frac{\norm{ V^k - X}^2}{2 \sigma^2}$, penalizes the difference between the individual reconstruction, $V^k$, and the consensus solution, $X$. $\sigma$ controls the convergence rate and the best $\sigma$ for convergence is determined experimentally. In every iteration of the CE method, each node evaluates Eqn.~(\ref{eqn:CE}), computes the system matrix for the $k^{th}$ forward model in Eqn.~(\ref{eqn:DS-FFS-CT}), and produces an individual solution, $V^k$. Then, individual reconstructions from all nodes are fused together for an updated consensus solution, $X$. If the new consensus solution, $X$, is different from individual reconstruction, $V^k$, then iterations repeat until $X= V^k$, where $k$ is from 1 to $K$. Otherwise, the consensus solution $X$ is the global minimum to Eqn.~(\ref{eqn:DS-FFS-CT}). Since the CE Method is not a theoretical contribution of this paper but a new application to DS-FFS CT reconstruction, we succinctly summarize the CE Method and its fusing operation in the following framework:
\begin{enumerate}
   \item For each forward model $k$ from 1 to $K$, we introduce a variable $U^k$ as an input change to the proximal map function and is initialized to equal $V^k$. In addition, we introduce $W^k$, which is a temporary copy of $V^k$.
   \item While individual reconstructions $V^1$, $V^2 ,$ $\dots$ $,V^K$ are not equal, we repeatedly do the following steps to each proximal map function and individual reconstruction:
   
   \begin{enumerate}
\item Compute individual reconstruction, $V^k \gets \argmin_{V^k} F_k(X+ U^k)$. \label{step:comp_individual}
\item Store a copy of $W^k$ as $(W^{'})^k$. \label{step:copy_w}
\item Compute $W^k \gets 2V^k - X - U^k$.
\item Update $W^k \gets \rho W^k + (1- \rho) (W^{'})^k$, where $\rho$ is a convergence parameter and is chosen to be between 0 and 1.
\item $X \gets \left(\sum_{k=1}^K W^k \right) / K$, so that the consensus solution $X$ is updated to be the arithmetic mean of $W^1$, $W^2 ,$ $\dots , $ $W^K$.
\item The proximal map function input change, $U^k$, is updated to be $U^k \gets X - W^k$. \label{step:update_change}
  \end{enumerate}
\end{enumerate}
In the above framework, step~\ref{step:comp_individual} updates each individual reconstruction by computing their proximal map functions. Steps~\ref{step:copy_w} to~\ref{step:update_change} fuse the individual reconstructions into a consensus solution, $X$. If $X$ does not equal to all individual reconstructions, the framework repeats.

\section{Experiment Setup}
\label{sec:experiment-setup}
We acquired data from a dual source Siemens Somatom Force CT scanner to assess the performance of the algorithms. The scanner at its default focal spot location has a 595 mm source-to-isocenter distance ($r_{so} =$ 595 mm), and a 1085.6 mm source-to-detector distance ($r_{sd} =$ 1085.6 mm). Detector sensor units are formed on an arc concentric to the X-ray source. At the single source mode, the CT detector has 96 rows and 920 channels, with a detector row spacing of 1.094 mm and a channel spacing of 0.054\textdegree. At the dual source mode, one detector of the scanner has 96 rows and 920 channels ($M_r$ = 96, $M^1_c$ = 920), and the other detector has a smaller field-of-view with 96 rows and 640 channels ($M^2_c$ = 640). In addition, the two detectors have a z direction translation offset of 0.88 mm ($\Delta z_s^{1,2}$ = 0.88 mm), and an angular offset of 95\textdegree in the X-Y plane ($\Delta \beta^{1,2}$ = 95\textdegree). Each source-detector pair has 9 possible focal spots and a CT scan can use any number of focal spots, depending on the scan protocol. In this paper, the exact displacement at each focal spot location, $( \Delta_u , \Delta_v , \Delta_z)$, is not disclosed as such information is confidential and protected by Siemens Healthineers.

\begin{figure}[t]
\centering
\begin{subfigure}{0.49\linewidth}
    \includegraphics[width=\linewidth]{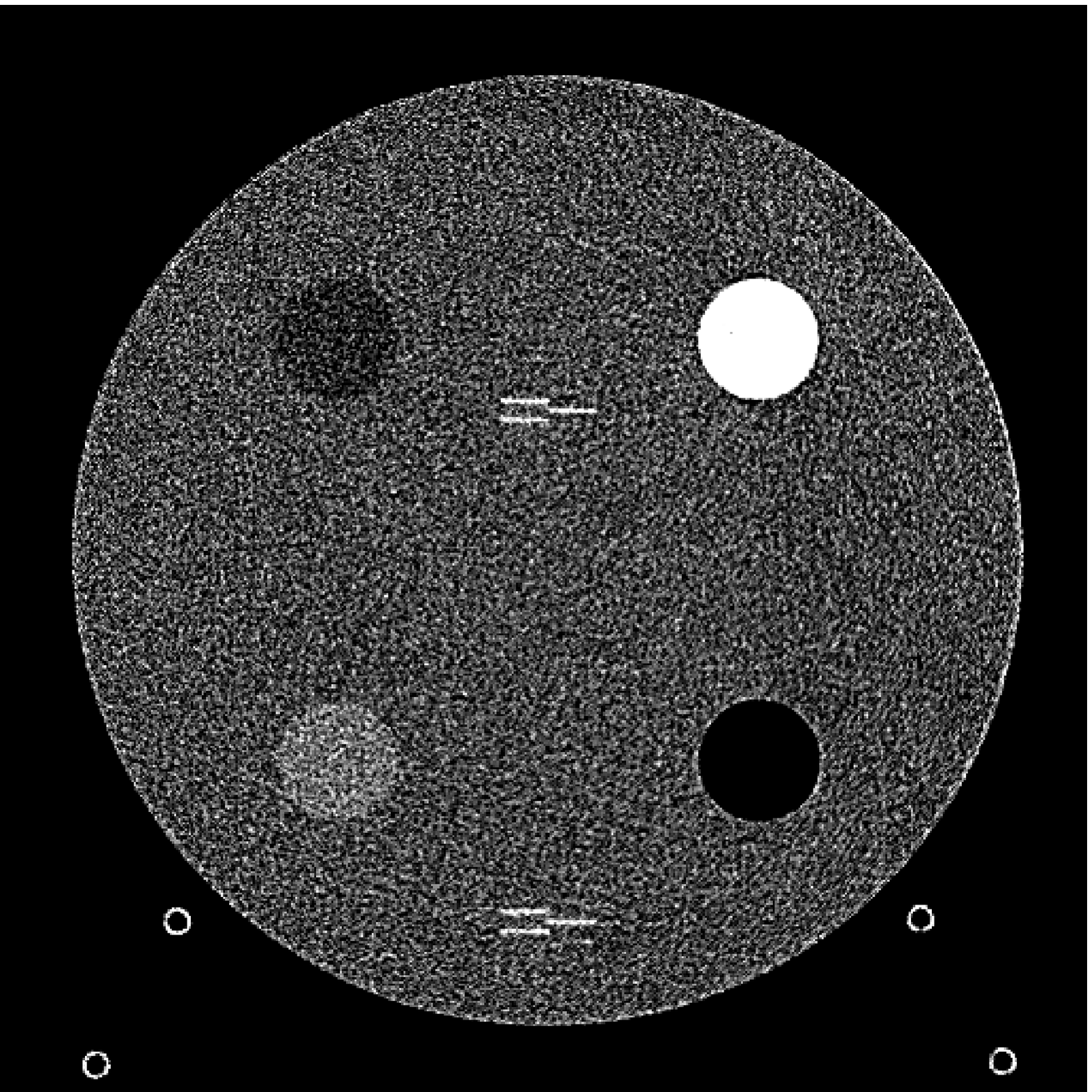}
    \caption{}
\end{subfigure}
\hfill
\begin{subfigure}{0.49\linewidth}
    \includegraphics[width=\linewidth]{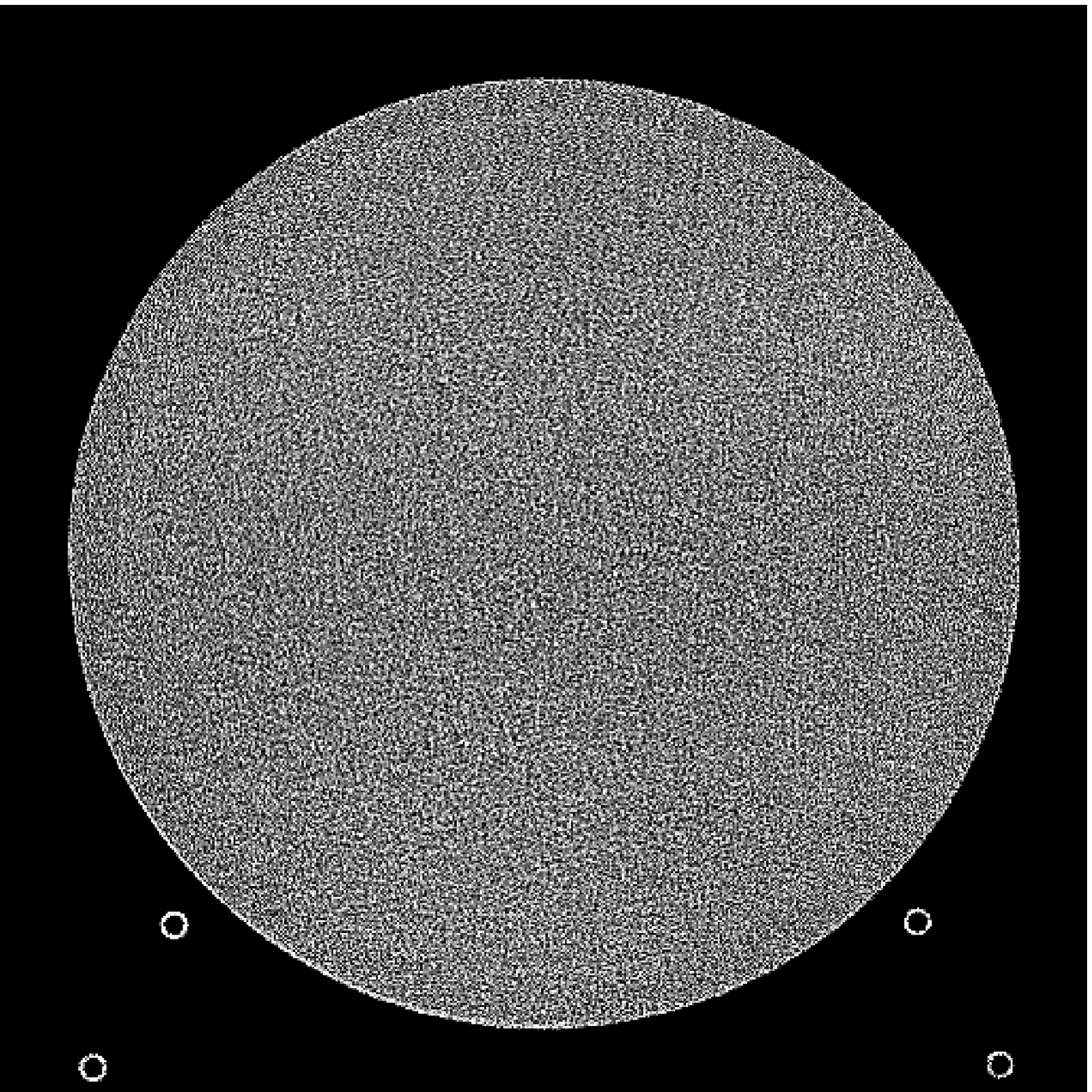}    
   \caption{}
\end{subfigure}
\caption{(a) The ACR phantom module 1 with 4 different inserts for CT number fidelity testing. In addition, two ramps of wires are visible near the phantom center. The image display window center is 150 HU and window width is 700 HU. (b) The phantom module 3 with a uniform water-equivalent radiodensity of 0 HU. The display window center is 0 HU and window width is 400 HU.}
   \label{fig:experiment-setup} 
  \vspace{1em}
\end{figure}

To evaluate image quality, we first tested the performance of JENG on a standard CT phantom, CT ACR 464 phantom~\cite{ACR-instruction}. Then we tested JENG on 5 clinical thoracic datasets and 3 abdominal datasets.  The ACR 464 phantom contains four modules in total, with each module 40 mm in depth and 200 mm in diameter. The first module has 4 different inserts to test CT number fidelity. In addition, the module contains a series of wires for cross-plane resolution evaluation, shown as white horizontal bars near the center in Fig.~\ref{fig:experiment-setup}(a), and are visible in 0.5 mm z-axis increments. The second module tests low contrast resolution, but is not used in this paper. The third module, shown in Fig.~\ref{fig:experiment-setup}(b), is a uniform cylinder of water-equivalent material of 0 Hounsfield units (HU), and we used this module to quantitatively measure image noise profile and in-plane resolution. The fourth module consists of resolution bars of various spatial frequencies for analysis on high contrast resolution. To scan the phantom, the scanner setup used dual sources with 2 focal spots at each source. In addition, the projections were acquired using the same protocol for clinical thoracic scans with 100 KV, a nominal tube current of 718 mA and a high helical pitch of 2.8. For all experiments, we compared JENG against the state-of-the-art Siemens ADMIRE, reconstructed with a BL-64 soft tissue sharp kernel, and we define ADMIRE as the clinical standard hybrid IR method for the rest of the paper. 

\begin{figure}
\begin{tikzpicture}[spy using outlines={circle,red,magnification=2,size=2.5cm, connect spies}]
\node {\includegraphics[width=.49\linewidth]{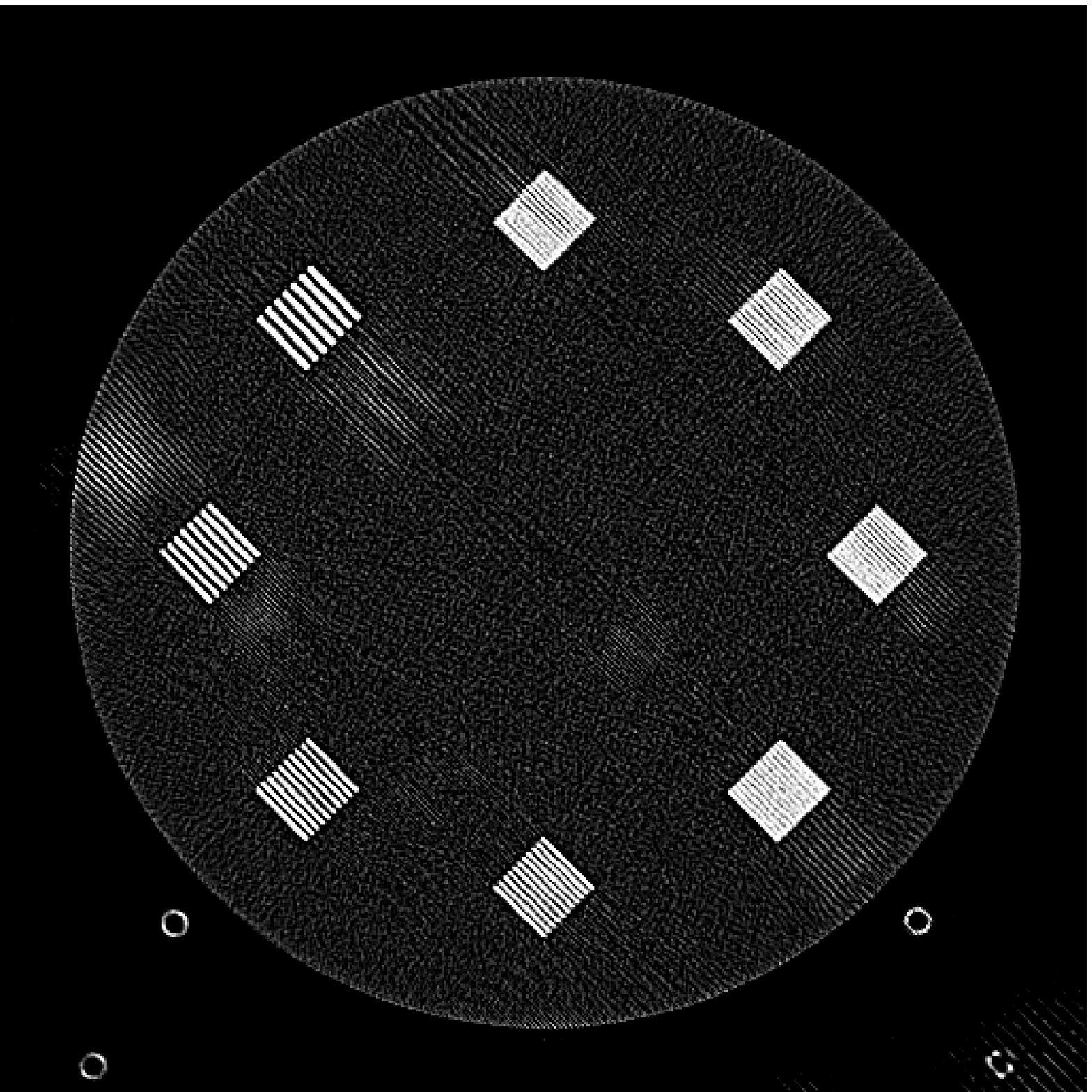}};
\spy on (1.8,-2.0) in node [left] at (4.8,-3.55);
\node at (-0.8,-4.7) {(a) Clinical Standard (L1 Denoising)};
\end{tikzpicture}
\begin{tikzpicture}[spy using outlines={circle,red,magnification=2,size=2.5cm, connect spies}]
\node {\includegraphics[width=.49\linewidth]{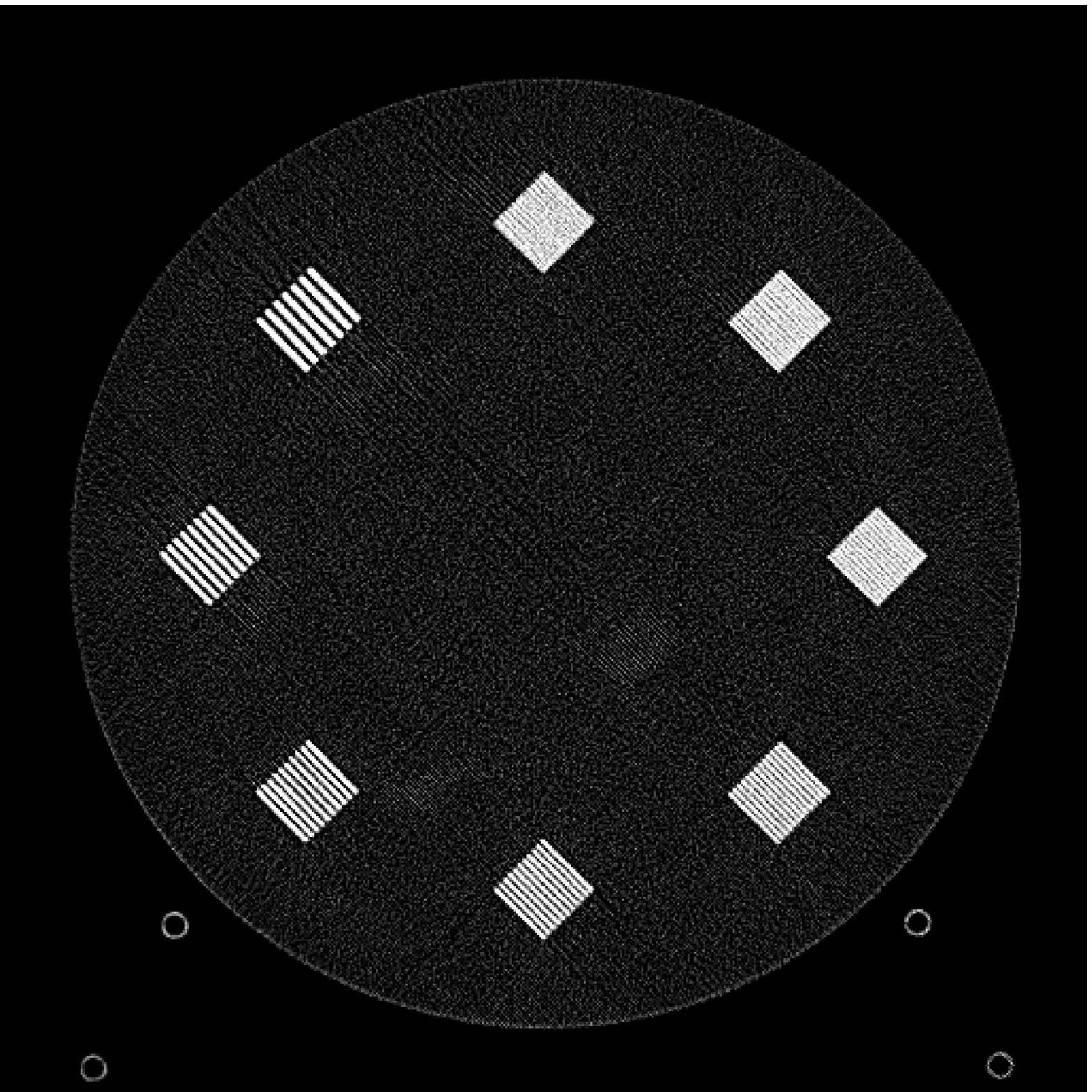}};
\spy on (1.8,-2.0) in node [left] at (4.8,-3.55);
\node at (-0.8,-4.7) {(b) JENG (L1 Denoising)};
\end{tikzpicture}
\begin{tikzpicture}[spy using outlines={circle,red,magnification=2,size=2.5cm, connect spies}]
\node {\includegraphics[width=.49\linewidth]{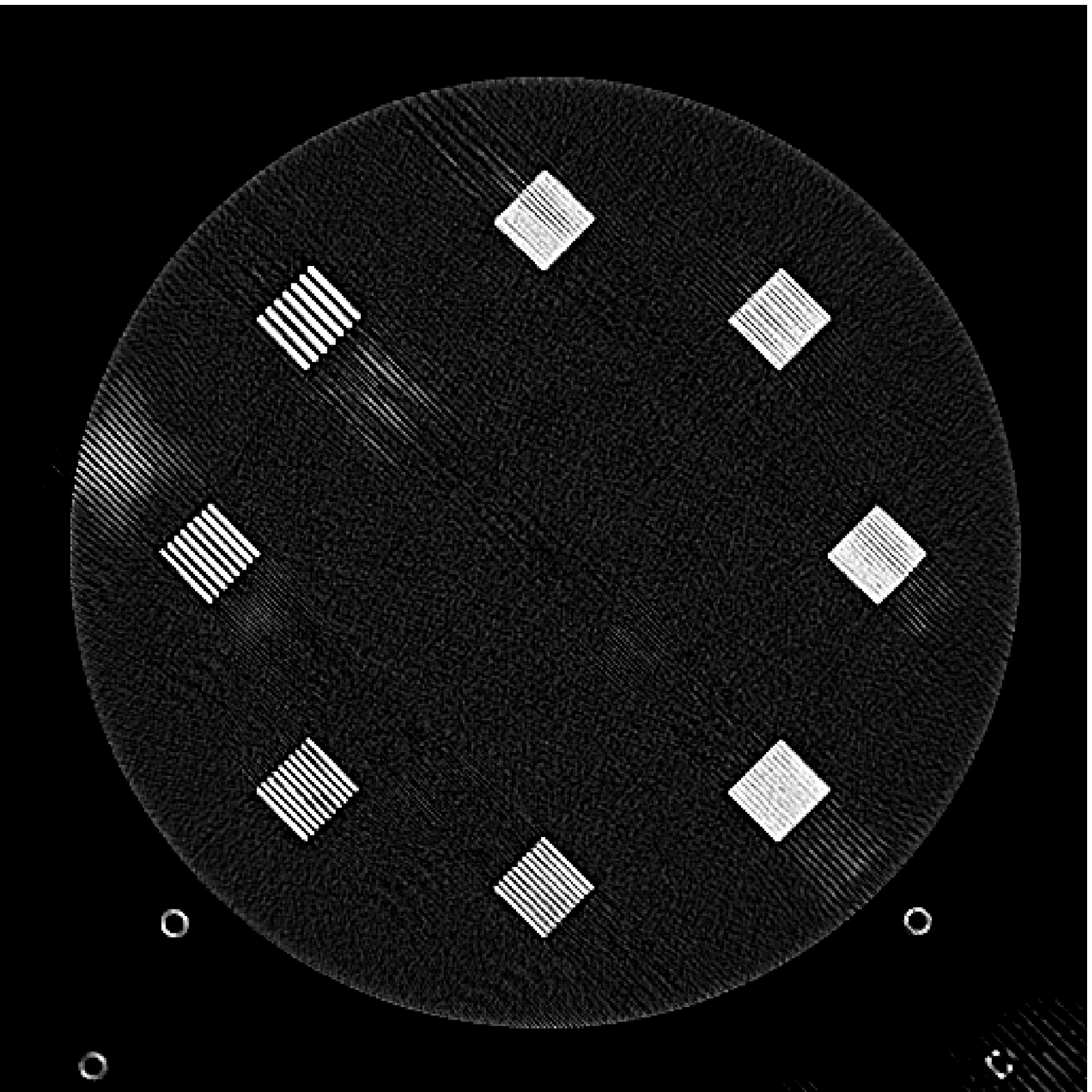}};
\spy on (1.8,-2.0) in node [left] at (4.8,-3.55);
\node at (-0.8,-4.7) {(c) Clinical Standard (L3 Denoising)};
\end{tikzpicture}
\begin{tikzpicture}[spy using outlines={circle,red,magnification=2,size=2.5cm, connect spies}]
\node {\includegraphics[width=.49\linewidth]{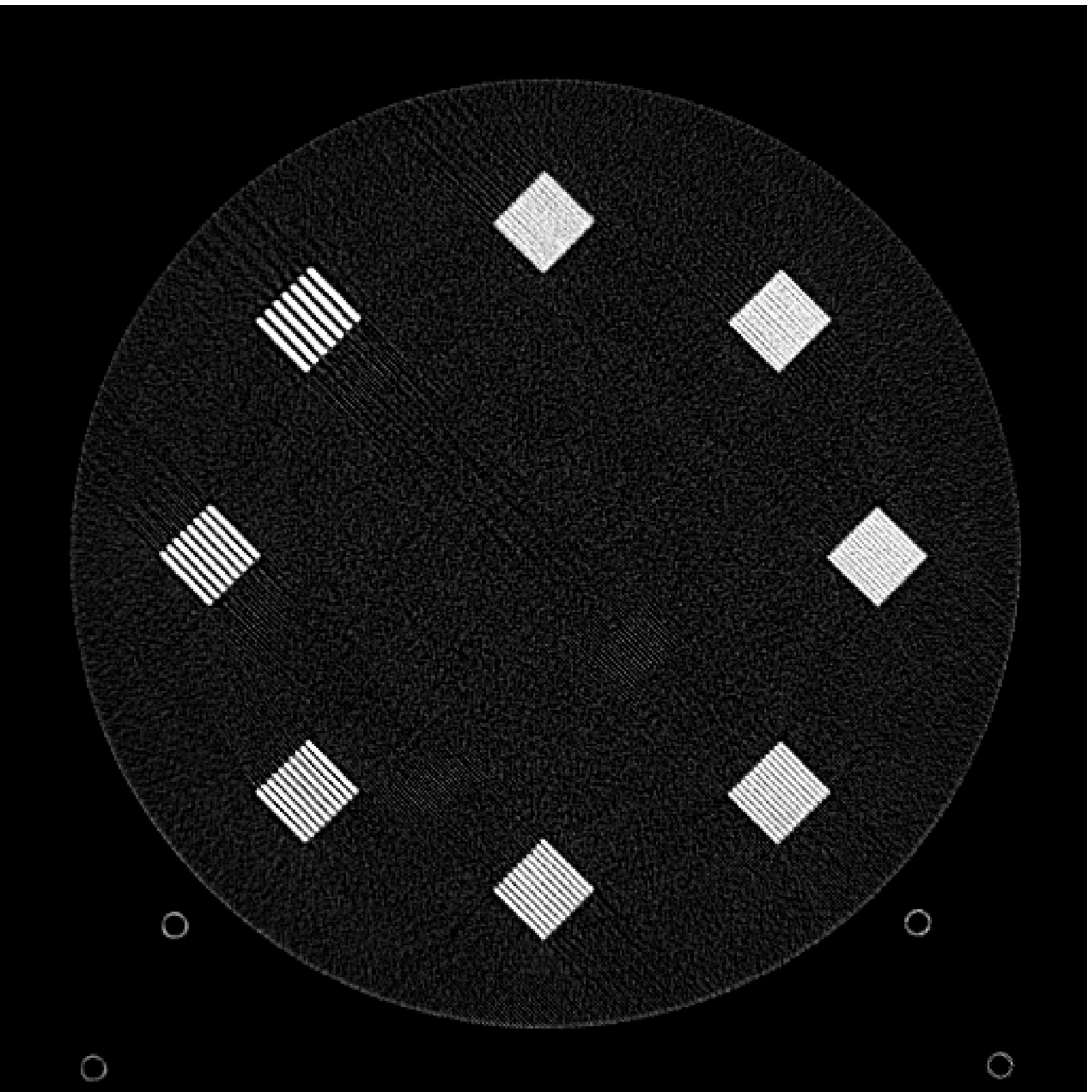}};
\spy on (1.8,-2.0) in node [left] at (4.8,-3.55);
\node at (-0.8,-4.7) {(d) JENG (L3 Denoising)};
\end{tikzpicture}
\caption{Clinical Standard hybrid IR vs JENG performance, with a display window center of 650 HU and a window width of 1500 HU. The spatial frequencies for bar patterns from top going clock-wise are 1.2, 1.0, 0.9, 0.8, 0.7, 0.6, 0.5, 0.4 mm$^{-1}$. (a) The clinical Standard hybrid IR at L1 denoising strength using a soft tissue high contrast kernel. Note that significant  aliasing streaking artifacts are present near the bar patterns. (b) JENG at a comparable L1 noise level but with much clearer bar pattern and fewer artifacts than the clinical standard hybrid IR. (c) The clinical standard hybrid IR at a stronger denoising strength of L3. (d) JENG at a comparable L3 noise level.}
\label{fig:ACR-high-resolution}
\end{figure}

The first experiment we performed was a visual comparison of in-plane spatial resolution between JENG and the clinical standard hybrid IR. We used the ACR phantom module 4 for this evaluation, which has 8 resolution bars of various spatial frequencies from 0.4 mm$^{-1}$ to 1.2 mm$^{-1}$. To obtain
a fair comparison, we matched the image noise variance in the uniform regions of JENG and the clinical standard hybrid IR and studied their in-plane spatial resolution and undersampling streaking artifacts. In addition, we performed two sets of experiments. The first set of experiments matched their image noise variance at the L1 denoising strength of the clinical standard method with a noise variance of 33926 in the uniform regions. The second set of experiments matched  their image noise variance at stronger L3 denoising strength of the clinical standard method with a noise variance of 12988 in the uniform regions.

The resolution bar visual comparison study, however, can be biased by observer subjectivity and may provide little information for spatial resolution beyond a limiting value.  Therefore, we also quantitatively evaluated the Task-Based Modulation Transfer Function (MTF\textsubscript{task}) of JENG and the clinical standard hybrid IR for a more complete analysis on in-plane resolution, using the edge of the uniform water-equivalent material phantom in module 3 as shown in  Fig.~\ref{fig:experiment-setup}(b). In summary, our MTF\textsubscript{task} analysis was measured with the water-equivalent phantom and averaged all transaxial images in module 3 into a 2D image. Then the MTF\textsubscript{task} analysis computed the oversampled edge-spread-function for the generated 2D image, differentiated and Fourier transformed the edge-spread-function to the frequency domain~\cite{Friedman13,Samei12}. The MTF\textsubscript{task} is then the absolute value of the Fourier Transform result and the source code of our MTF\textsubscript{task} computations can be downloaded from citation~\cite{Friedman20}.

For a more complete image analysis, we also measured the Noise Power Spectrum (NPS) and we visually compared the cross-plane spatial resolution of JENG and the clinical standard hybrid IR. The NPS computations followed the same procedures as in citation~\cite{Friedman13} with source code from citation~\cite{Friedman20}. In summary, we selected multiple regions of interest in module 3 and all regions were squares of the same size and had an average radiodensity of 0 HU. In addition, neighboring regions of interest overlapped with each other. Then, we performed a Fourier Transform on each region of interest and the final NPS value equals to the ensemble average of the squared Fourier Transform.
For cross-plane resolution qualitative evaluation, we visually compared the wire pattern image sharpness between the two algorithms for the series of wires of module 1 from sagittal view. For qualitative evaluations, we reconstructed JENG with a resolution no worse than the clinical standard hybrid IR and compared image noise and artifacts between the two algorithms.

Finally, we tested the qualitative clinical results of JENG on 5 thoracic and 3 abdominal scan datasets in terms of spatial resolution, artifacts, image noise and low-contrast detectability. All clinical scans were retrospective and were acquired during routine clinical practice at a major children's hospital in the United States. All thoracic scans used the same settings as in the ACR phantom scans, except that the tube current was modulated differently for each patient based on their body thickness, weight, and age. The abdominal scans, however, had a reduced helical pitch at 0.6 to improve low contrast lesion detectability. For all clinical images in this paper, the clinical standard hybrid IR used a soft tissue high-contrast kernel for reconstructions and we compared the image noise and artifacts of JENG and the clinical standard hybrid IR after we matched their spatial resolution.

\section{Results}
\label{sec:results}
\subsection{ACR Phantom Study}
\label{subsec:acr-result}

\begin{wrapfigure}{R}{0.41\textwidth}
\begin{center}
\includegraphics[width=.40\textwidth]{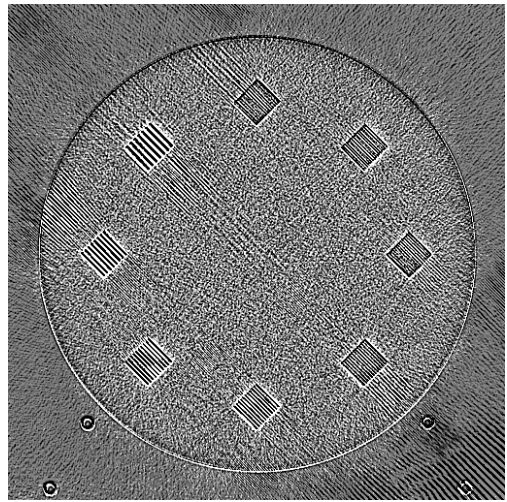}
\end{center}
\caption{Difference image between the clinical standard hybrid IR and JENG at a denoising strength of L1.}
\label{fig:error-ACR} 
\end{wrapfigure}

Fig.~\ref{fig:ACR-high-resolution} is an example image for resolution bars and the spatial frequencies for the bar patterns from top going clockwise are 1.2, 1.0, 0.9, 0.8, 0.7, 0.6, 0.5 and 0.4 mm$^{-1}$. Fig.~\ref{fig:ACR-high-resolution}(a) is the resolution bars reconstructed by the clinical standard hybrid IR at L1 denoising strength. Fig.~\ref{fig:ACR-high-resolution}(b) is JENG reconstructed at an image noise variance comparable to the L1 denoising. Fig.~\ref{fig:ACR-high-resolution}(c) is the clinical standard hybrid IR at a stronger L3 denoising strength and we can observe that the result at L3 denoising leads to less image noise than the result at L1 denoising in Fig.~\ref{fig:ACR-high-resolution}(a). Fig.~\ref{fig:ACR-high-resolution}(d) is JENG at an image noise variance comparable to the L3 denoising. To help readers better see the image quality difference between the clinical standard hybrid IR and JENG, Fig.~\ref{fig:error-ACR} is the difference image between the two algorithms at L1 denoising.
A noticeable difference between the clinical standard hybrid IR and JENG is that the clinical standard method in Figs.~\ref{fig:ACR-high-resolution}(a) and (c) have strong undersampling aliasing artifacts near the phantom periphery, which show a pattern of high density streakings and the streakings point along the direction of X-rays. In addition, the magnified sub-figures show that the bar pattern at 0.8 cycles/mm is unresolved with blurry details. In contrast, JENG in Figs.~\ref{fig:ACR-high-resolution}(b) and (d) effectively decimates the undersampling artifacts and the bar pattern in the magnified sub-figures of JENG is completely resolved with clearer details. 

For the clinical standard hybrid IR, a possible cause for its loss of image resolution and the presence of aliasing artifacts in the image periphery can be explained by the Nyquist-Shannon sampling theorem. Nyquist-Shannon sampling theorem concludes that the discrete projection sampling rate for an application that requires Fourier Transform and data interpolation must be sufficiently high to avoid alias in the frequency domain and capture all the needed information in the continuous image domain. Given that the clinical standard hybrid IR involves Fourier Transform and data interpolation operations, the low projection sampling rate at a high helical pitch of 2.8 might lead to aliasing artifacts and a loss of spatial resolution for the clinical standard method. In contrast, JENG has no Fourier Transform or data interpolation operations and is completely based on linear algebra and acquisition physics modeling. Therefore, JENG is not limited by Shannon-Nyquist Theorem and its images are less susceptible to aliasing artifacts and show clearer bar patterns.

\begin{figure}
\centering
\begin{subfigure}{0.49\linewidth}
\includegraphics[width=\linewidth]{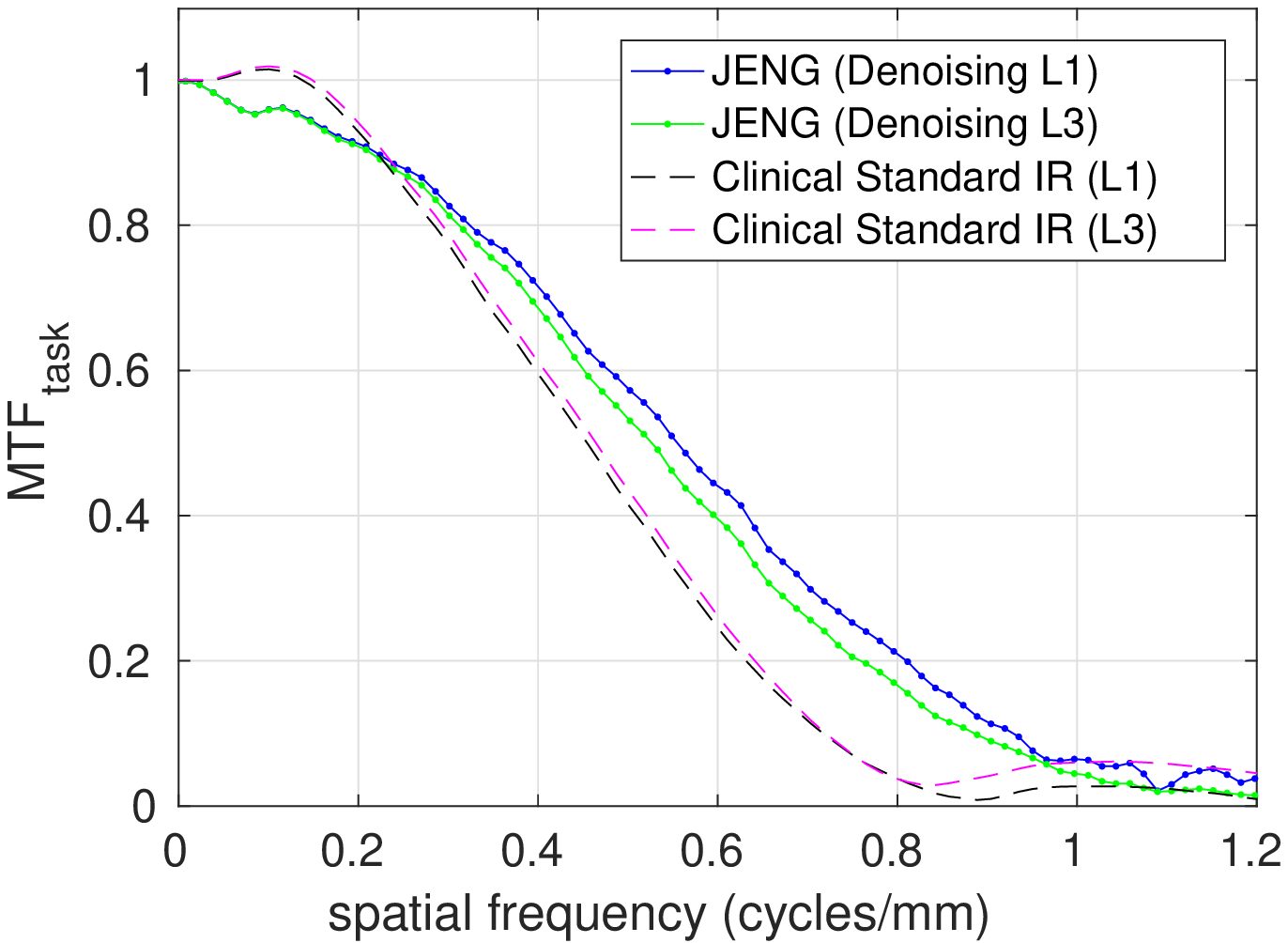}
\caption{MTF Plot}
\end{subfigure}
\begin{subfigure}{0.49\linewidth}
\includegraphics[width=\linewidth]{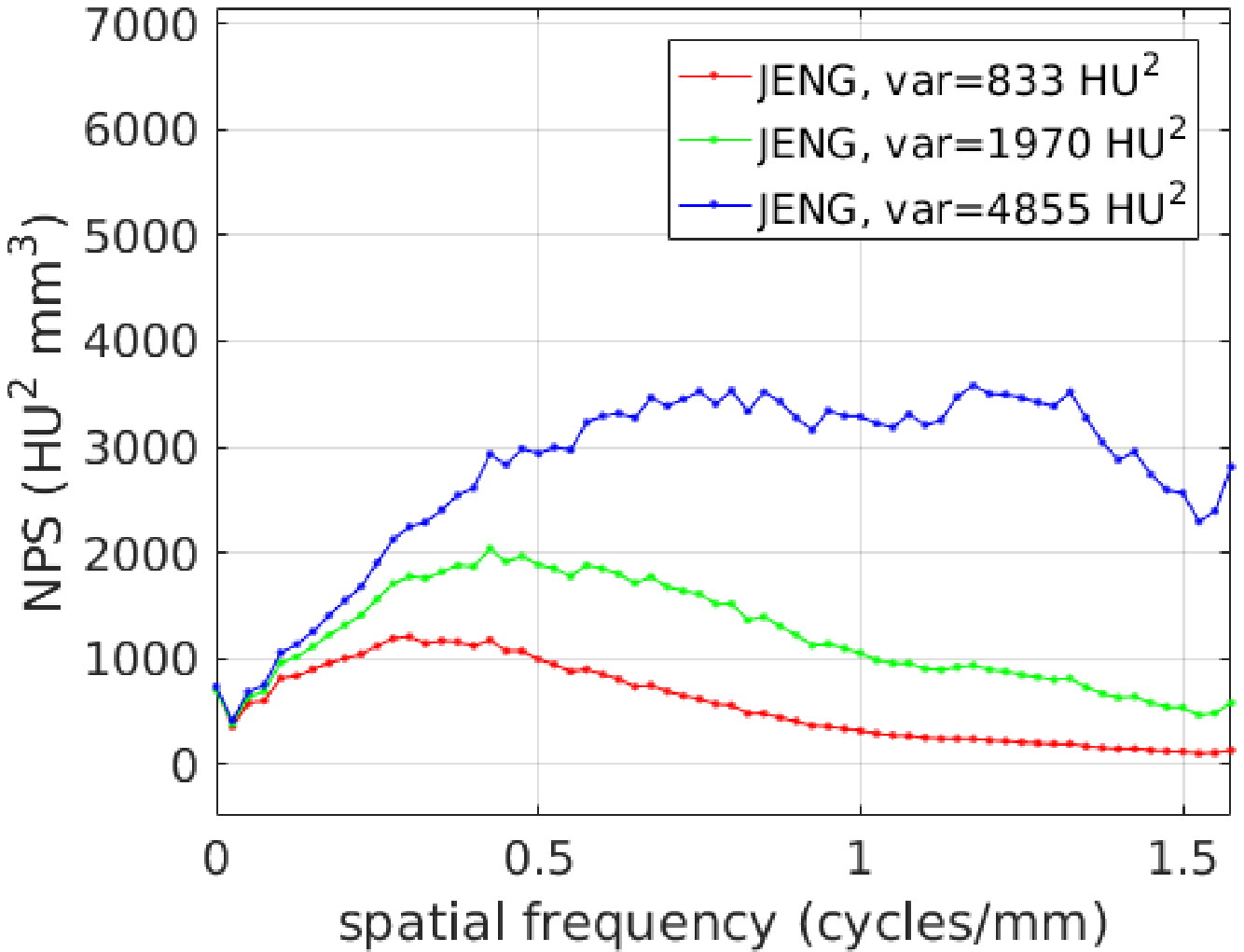}
\caption{NPS Plot}
\end{subfigure}
\caption{(a) The MTF\textsubscript{task} for the clinical standard hybrid IR and JENG, measured with a water-equivalent phantom. Note that JENG has a higher MTF\textsubscript{task} than the other at frequencies ranging from 0.3 mm$^{-1}$ to 1 mm$^{-1}$. (b) The 3D NPS for JENG at different noise levels. Note that a lower noise variance shifts JENG's NPS peak towards a lower spatial frequency. This phenomenon indicates that a strong denoising for JENG can alter image noise texture.}
\label{fig:MTF-NPS} 
  \vspace{1em}
\end{figure}

To corroborate with our visual assessment that JENG has a higher in-plane resolution than the clinical standard hybrid IR, we compare the two algorithms' image sharpness quantitatively through MTF\textsubscript{task} in Fig.~\ref{fig:MTF-NPS}(a), measured with a water-equivalent material phantom. We observe that the MTF\textsubscript{task} for the clinical standard hybrid IR drops to 0.1 at around 0.7 mm$^{-1}$, where an MTF\textsubscript{task} value of 0.1 is often considered the lowest contrast sensitivity for human visual observation. This observation aligns with our qualitative assessment in Fig.~\ref{fig:ACR-high-resolution}, which shows that bar patterns for the clinical standard hybrid IR becomes unintelligible at a spatial frequency near 0.7 mm$^{-1}$. In contrast, JENG has a higher MTF\textsubscript{task} than the clinical standard method at nearly all spatial frequencies ranging from 0.3 mm$^{-1}$ to 1 mm$^{-1}$. In addition, the MTF\textsubscript{task} value of JENG drops to 0.1 at around 1.0 mm$^{-1}$, suggesting that JENG can clearly show bar patterns until spatial frequency reaches 1 mm$^{-1}$. Therefore, both qualitative and quantitative assessments have an agreed conclusion that JENG has a higher in-plane spatial resolution than the clinical standard hybrid IR for the phantom study.

The image noise profile is another topic of interest. Fig.~\ref{fig:MTF-NPS}(b) is the NPS for JENG at three noise levels with image noise variances at 833 HU$^2$, 1970 HU$^2$ and 4855 HU$^2$. It is not surprising to observe that at all spatial frequencies JENG at higher noise variance has a larger NPS magnitude than that at lower noise variance. We also note that a lower noise variance shifts the NPS peak of JENG towards a lower spatial frequency, and this phenomenon indicates that a stronger denoising for JENG might alter the image noise texture.
Furthermore, we observe that at an extremely low spatial frequency below 0.1 mm$^{-1}$, JENG retains moderate noise with an NPS magnitude at 750 HU$^2$ and fail to denoise further.
To understand why JENG has limited success in denoising at a very low spatial frequency, we need to revisit JENG's prior model, $R(X)$, in Eqn.~(\ref{eqn:mbir}). As explained before, $R(X)$ is a local-neighbor Markov Random Field, and denoise each voxel based on the voxel's difference with its neighbors. The Markov Random Field prior model, however, is a low-pass filter. Therefore, $R(X)$ can suppress high-frequency noise well and preserve high-contrast image edges, but has limited success in low-frequency denoising and retains some very low frequency noise in the JENG images.

\begin{figure}
\centering
\begin{subfigure}{0.45\linewidth}    
    \includegraphics[angle=90,origin=c,trim=0.7cm 1.5cm 14.5cm 10cm,clip,width=\linewidth]{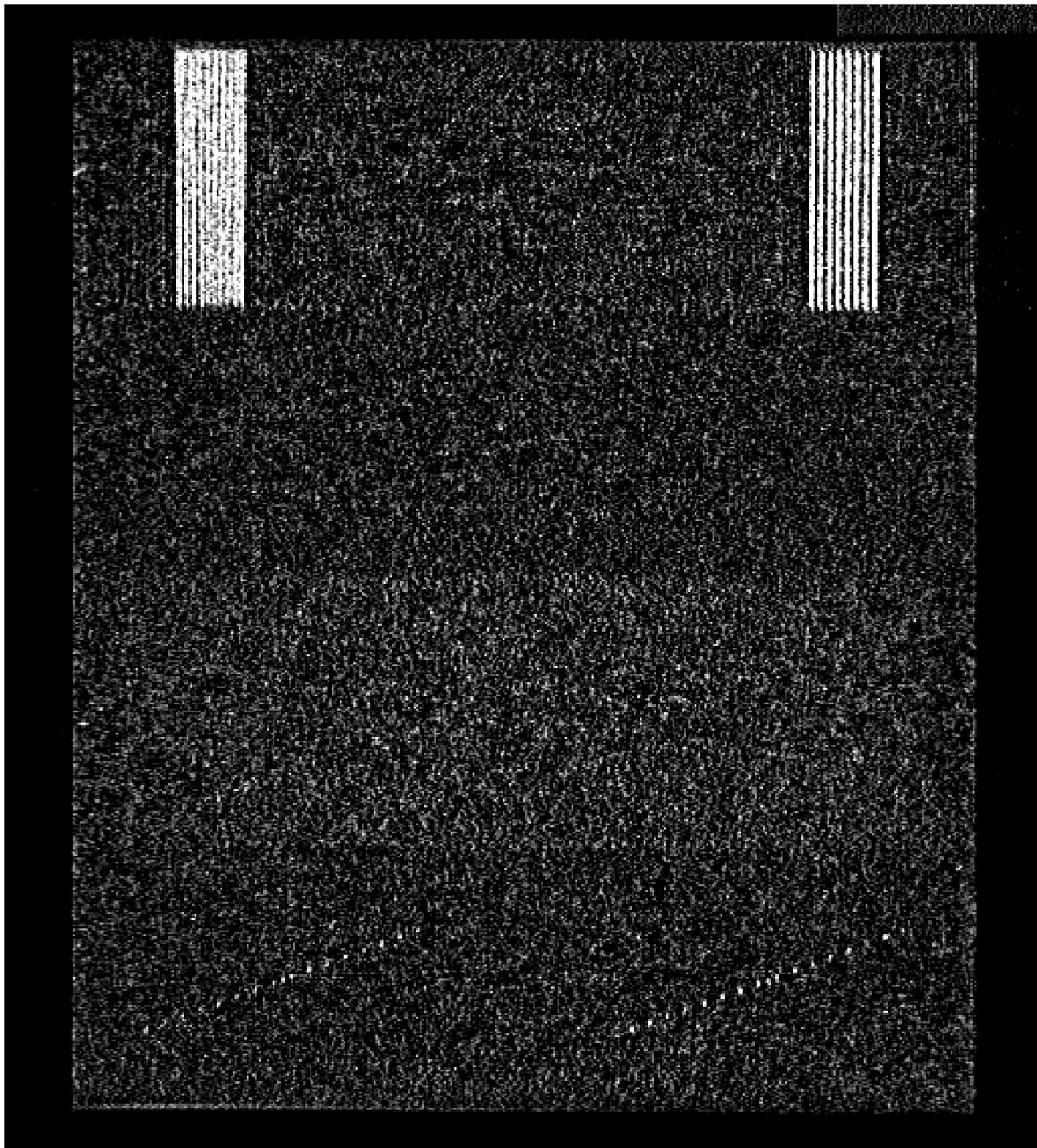}
    \vspace{-15mm}    
    \caption{Clinical Standard Hybrid IR}
\end{subfigure}
\begin{subfigure}{0.45\linewidth}    
    \includegraphics[angle=90,origin=c, trim=0.7cm 1.5cm 14.5cm 10cm,clip,width=\linewidth]{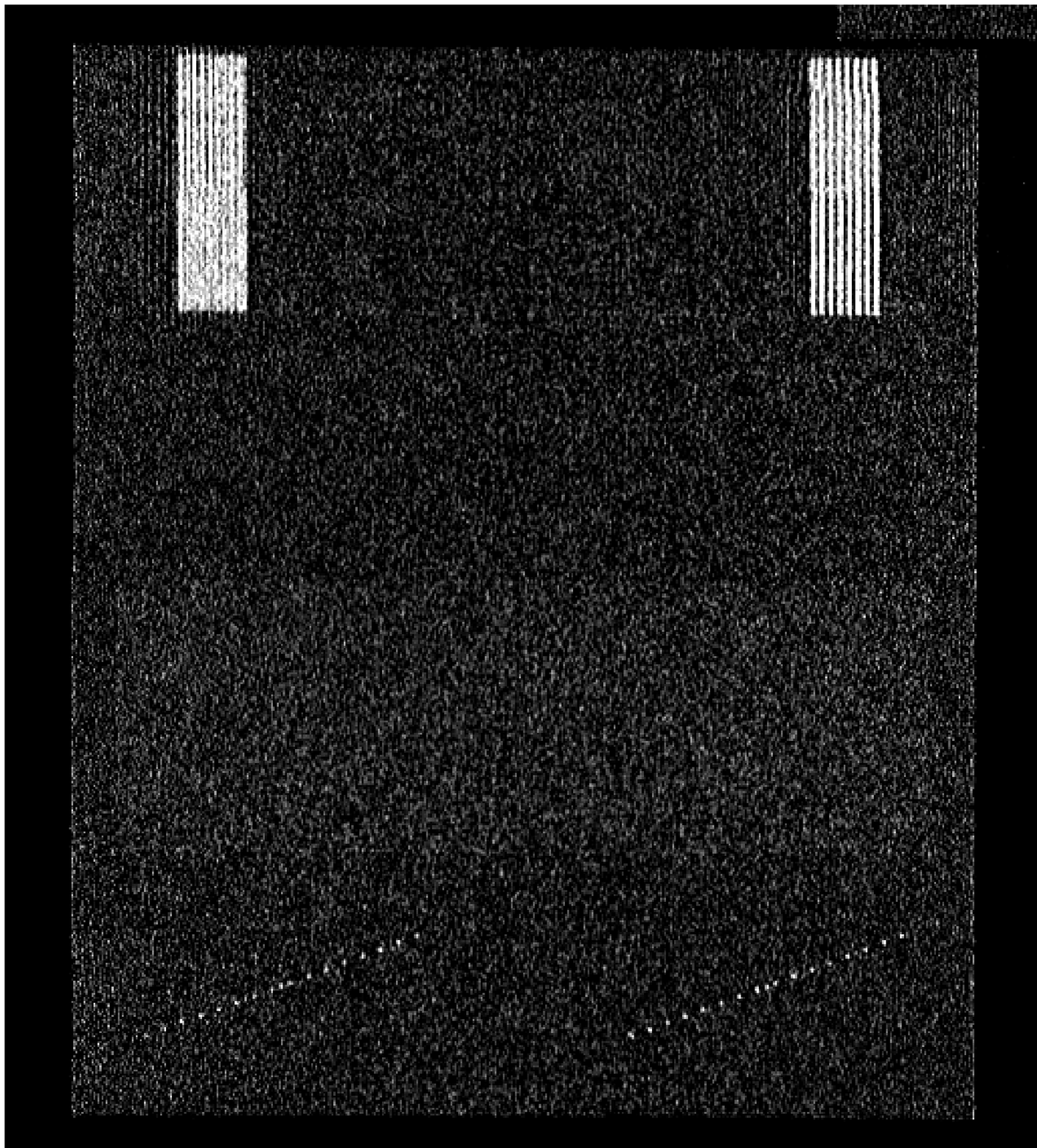}
    \vspace{-15mm}    
    \caption{JENG}
\end{subfigure}
\caption{Cross-plane spatial resolution comparison between the clinical standard hybrid IR and JENG. (a) The clinical standard hybrid IR, with a display window center at 650 HU and a window width of 1500 HU. Note that some wires are obscured in the image. (b) At a resolution no worse than the clinical standard hybrid IR, note that JENG better suppresses image noise and have fewer smear artifacts than the clinical standard hybrid IR.}
\vspace{3mm}    
\label{fig:ACR_cross-plane}
\end{figure}

 For many clinical applications, cross-plane resolution is equally important to in-plane resolution. To evaluate cross-plane resolution, we visually compared the clinical standard hybrid IR and JENG's wire series in module 1 from sagittal view with neighboring wires 0.5 mm apart from each other along the z axis. Fig.~\ref{fig:ACR_cross-plane}(a) is the wire series from cross-plane sagittal view, reconstructed by the clinical standard hybrid IR. We can observe that some of the wire series have blurry smears and the visibility of the wire series is impacted by significant image noise.  Fig.~\ref{fig:ACR_cross-plane}(b) is the wire series reconstructed by JENG with a resolution no worse than the clinical standard hybrid IR. We observe that the wire series are more visible due to less image noise. In addition, the wire series have fewer smearing artifacts than the clinical standard hybrid IR.

\begin{figure}
\begin{tikzpicture}[spy using outlines={circle,red,magnification=1.5,size=3cm, connect spies}]
\node {\includegraphics[width=.45\linewidth]{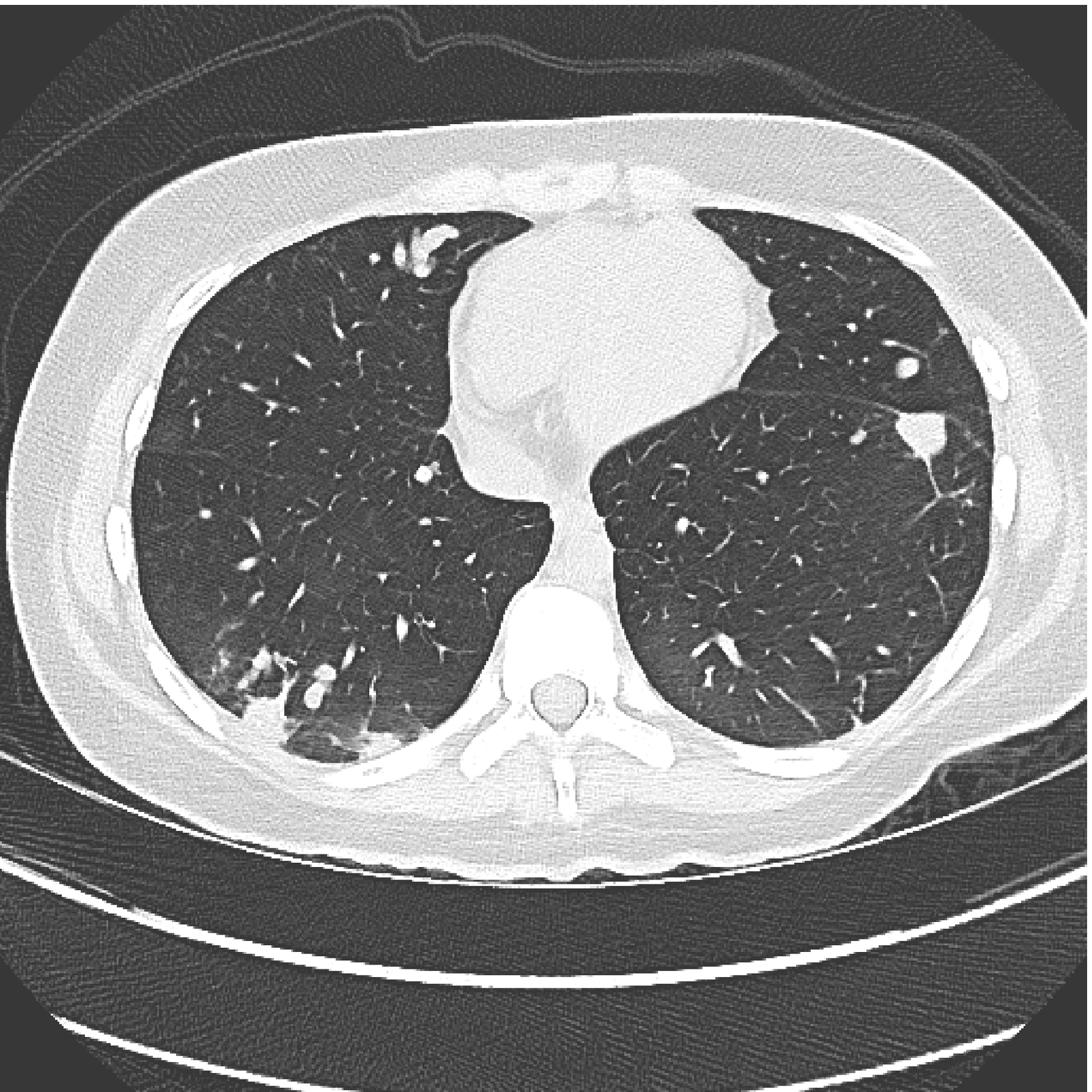}};
\spy on (-2.0,2.2) in node [left] at (-0.85,-3.75);
\node at (1.4,-4.4) {(a) Clinical Standard IR};
\end{tikzpicture}
\begin{tikzpicture}[spy using outlines={circle,red,magnification=1.5,size=3cm, connect spies}]
\node {\includegraphics[width=.45\linewidth]{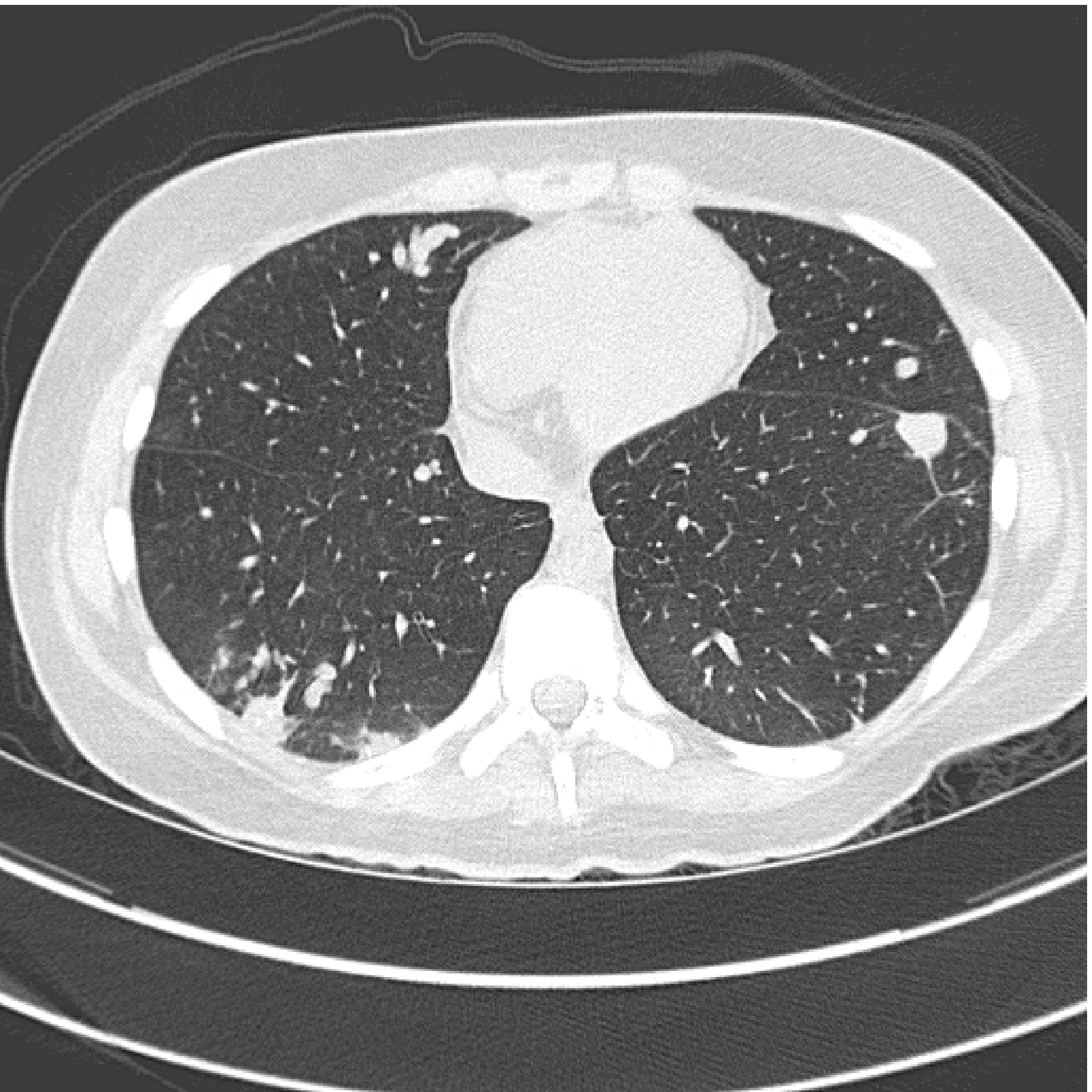}};
\spy on (-2.0,2.2) in node [left] at (-0.85,-3.75);
\node at (0.5,-4.4) {(b) JENG};
\end{tikzpicture}
\begin{tikzpicture}[spy using outlines={circle,red,magnification=1.5,size=3cm, connect spies}]
\node {\includegraphics[width=.45\linewidth]{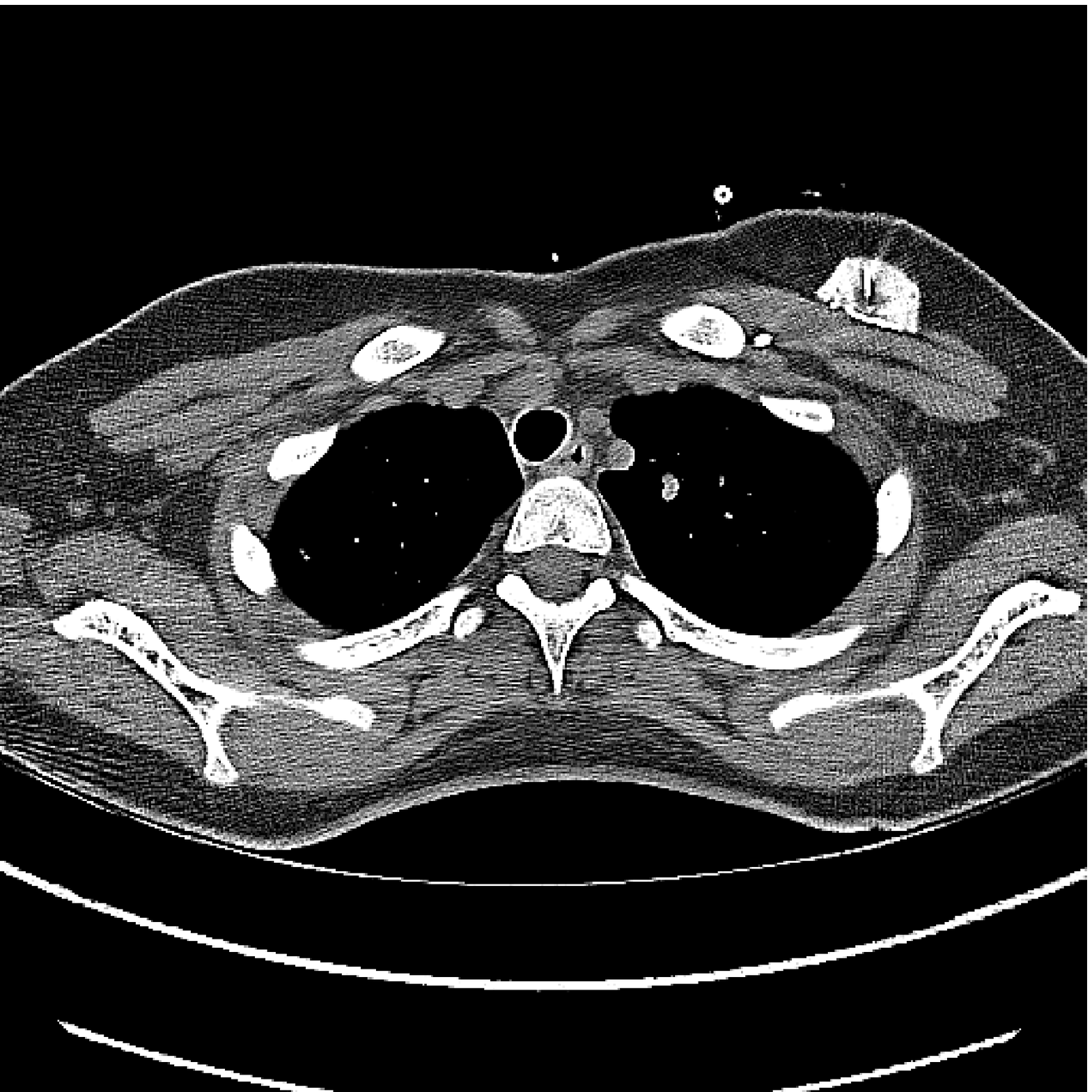}};
\spy on (-2.7,0.5) in node [left] at (4.85,-2.75);
\node at (-1.1,-4.4) {(c) Clinical Standard IR};
\end{tikzpicture}
\begin{tikzpicture}[spy using outlines={circle,red,magnification=1.5,size=3cm, connect spies}]
\node {\includegraphics[width=.45\linewidth]{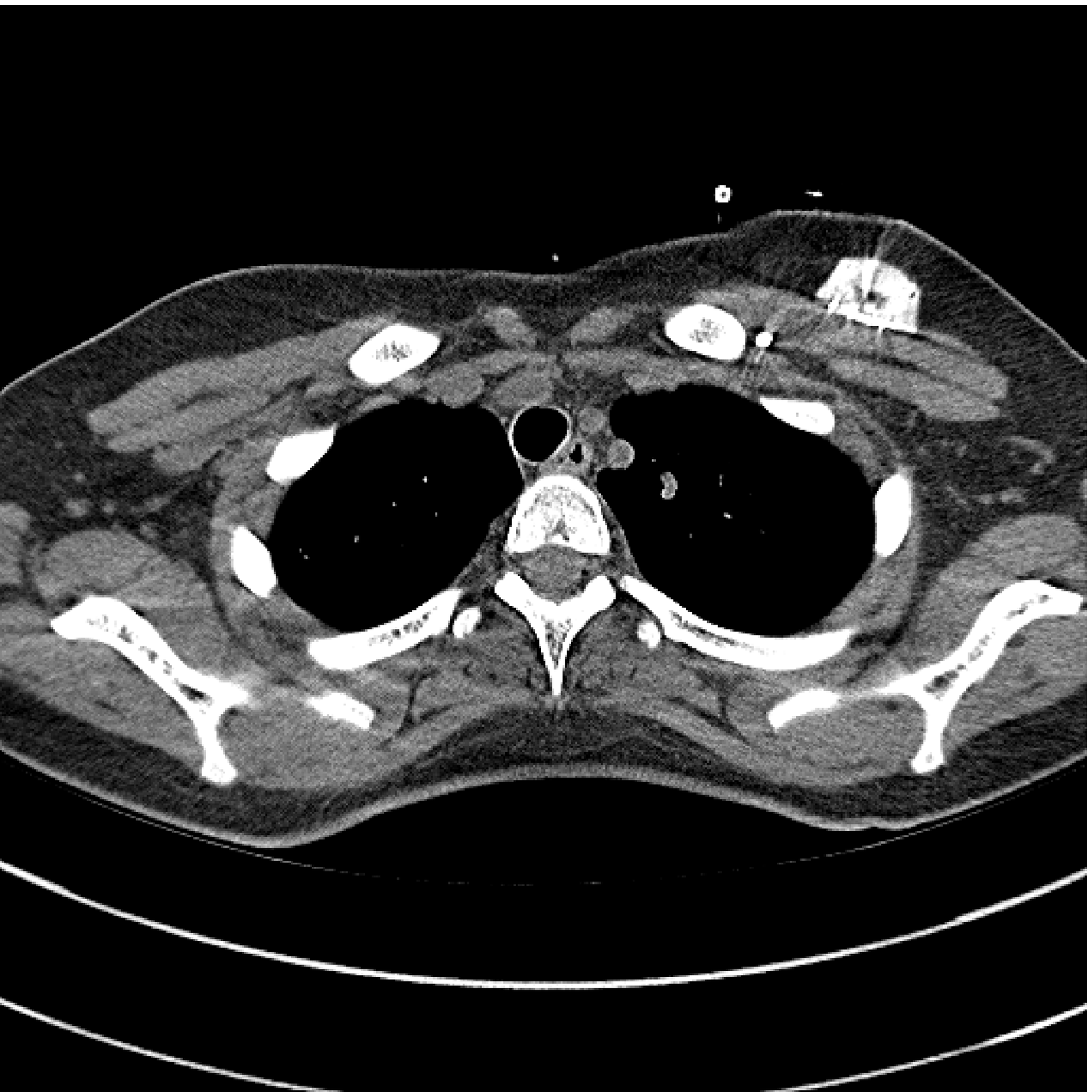}};
\spy on (-2.7,0.5) in node [left] at (4.85,-2.75);
\node at (-1.1,-4.4) {(d) JENG};
\end{tikzpicture}
\caption{Qualitative clinical results from a thoracic CT staging dataset for a 12-year-old with osteosarcoma with pulmonary metastases. The clinical standard hybrid IR is shown on the left and JENG is on the right. (a) The clinical standard hybrid IR in lung window with a window center of -600 HU and a window width of 1500 HU.  (b) JENG at a resolution comparable to the clinical standard, but with less noise and fewer artifacts. (c) The clinical standard hybrid IR in soft tissue window with a window center of 55 HU and a window width of 440 HU. A metastatic lung cancer nodule can be found 
in the left upper lobe. (d) JENG in soft tissue window at a comparable resolution, but with less noise and fewer artifacts. Note that JENG is not fully corrected for beam hardening artifacts.}
\label{fig:lung_clinical_images}
  \vspace{1em}
\end{figure}

\begin{figure}
\centering
\begin{subfigure}{0.40\linewidth}
\includegraphics[width=\linewidth]{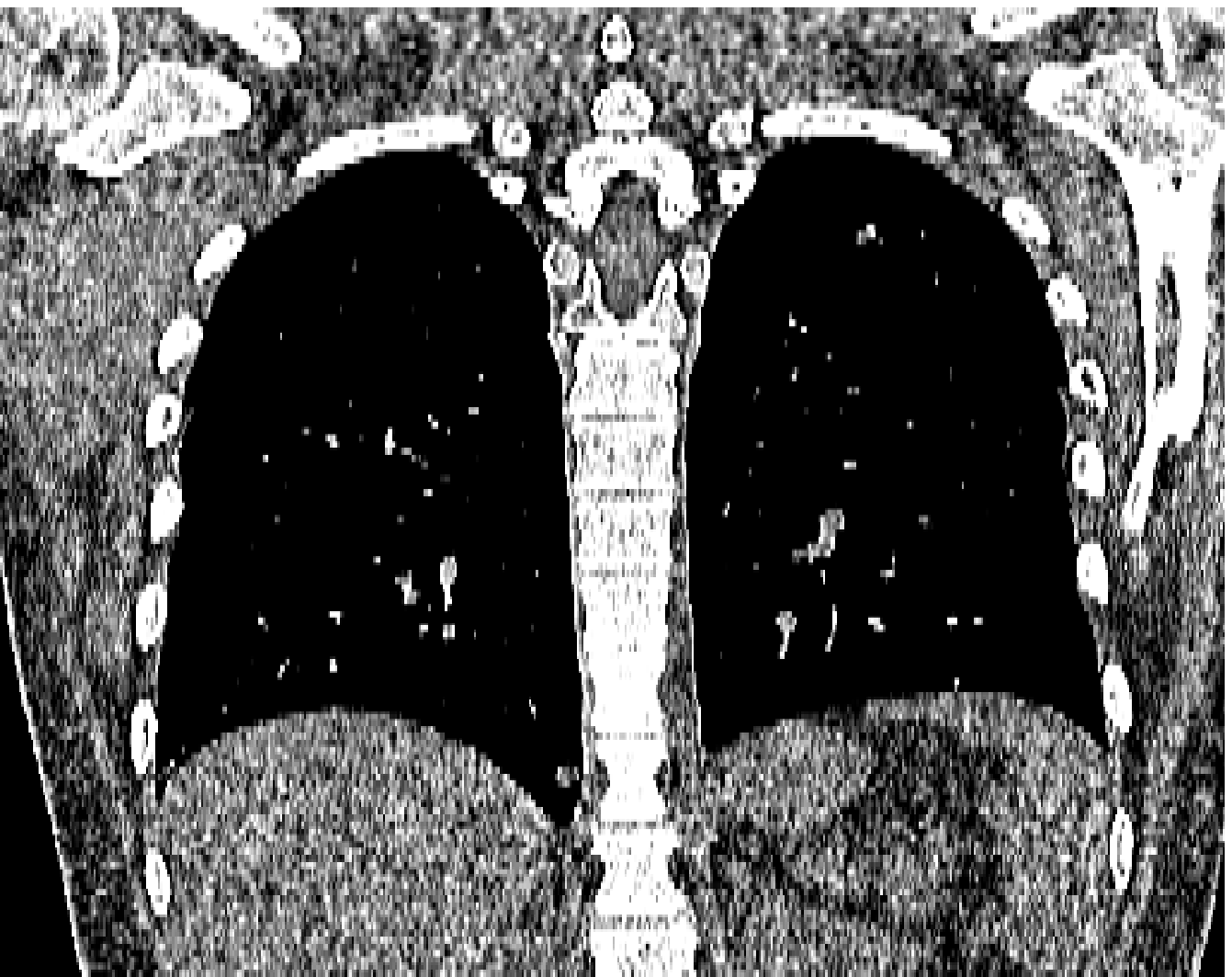}
\caption{Clinical Standard Hybrid IR}
\end{subfigure}
\begin{subfigure}{0.40\linewidth}
\includegraphics[width=\linewidth]{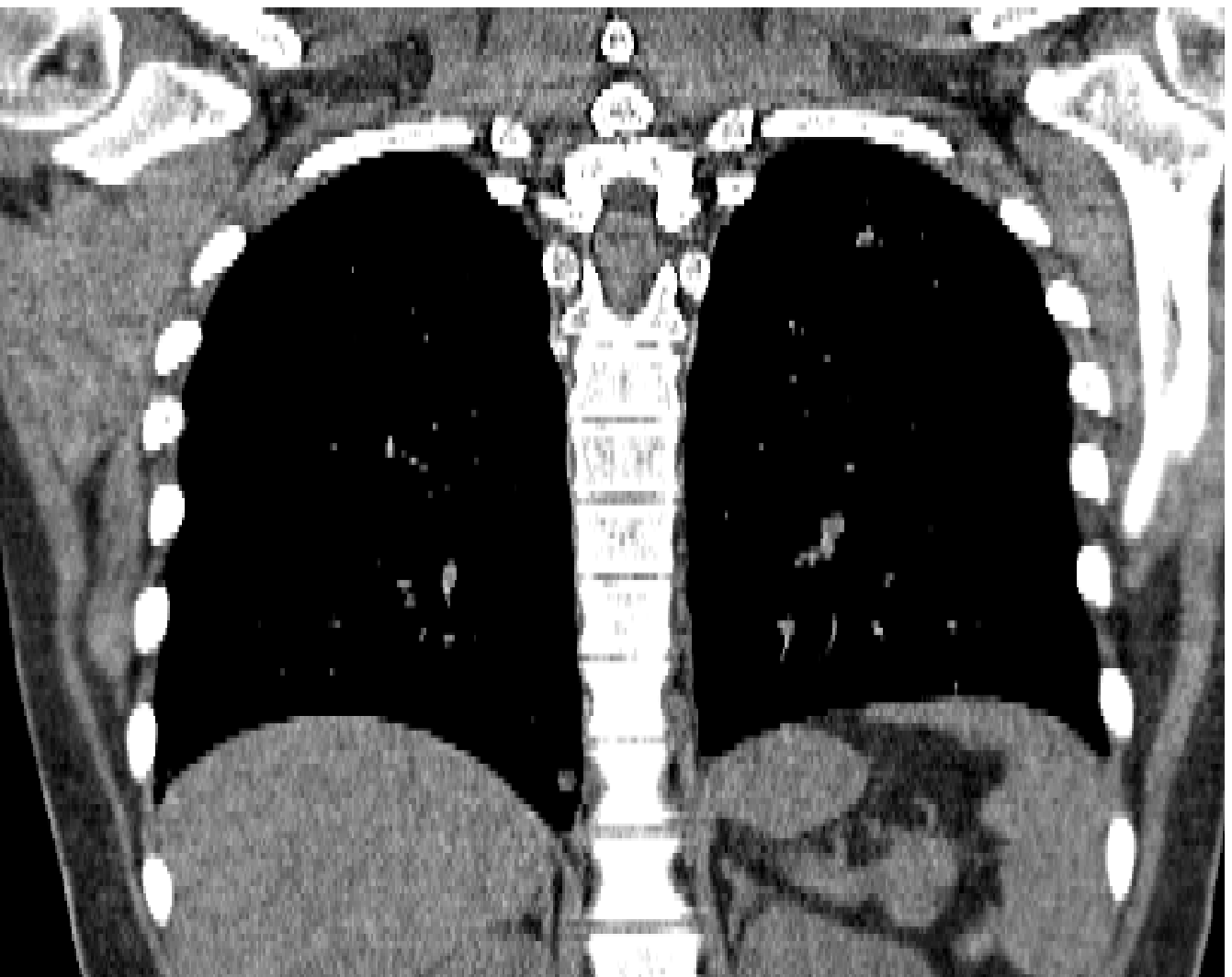}
\caption{JENG}
\end{subfigure}
\caption{An example cross-plane image from the same thoracic dataset as in Fig.~\ref{fig:lung_clinical_images}. (a) A coronal-view image slice of the clinical standard hybrid IR in soft tissue window. (b) JENG at a comparable resolution but with reduced image noise and artifacts.}
\label{Fig:coronal_clinical_images}
  \vspace{1em}
\end{figure}

\subsection{Clinical Cases}
\label{subsec:clinical-result}
None of the spatial resolution and artifact reduction advantages would hold unless JENG shows image quality improvement over the clinical standard method on patient datasets. To do so, we evaluated JENG on 5 thoracic and 3 abdominal CT scans and all scans used the same parameter settings as those for the ACR phantom scan, except that the tube current and the pitch is modulated individually for each patient. The exact experiment setup was discussed in Sec.~\ref{sec:experiment-setup}. For a fair image quality comparison, we matched the resolution of JENG and the clinical standard hybrid IR and studied their image noise and artifacts.

Figs.~\ref{fig:lung_clinical_images}(a) and (b) are an example thoracic scan image from a staging dataset for a 12-year-old with osteosarcoma with pulmonary metastases prior to a surgery. Fig.~\ref{fig:lung_clinical_images}(a) is the  clinical standard hybrid IR image reconstructed with a BL-64 soft tissue sharp kernel and display in the lung window. Aliasing streaking artifacts from undersampling are present almost everywhere in the image and artifacts are more pronounced near the image periphery. Fig.~\ref{fig:lung_clinical_images}(b) is the same image slice reconstructed by JENG at a comparable resolution. We notice that JENG decimates the aliasing streaking artifacts to a large extent, despite that the upper left lobe of the image still retain some undersampling artifacts. Fig.~\ref{fig:lung_clinical_images}(c) is another clinical standard hybrid IR image from the same dataset and is displayed in the soft tissue window. In this example image, a metastatic cancer nodule can be found in the image near the upper left lobe. Overall, the image quality is negatively influenced by its strong image noise and aliasing streaking artifacts. In contrast, JENG at a comparable resolution in Fig.~\ref{fig:lung_clinical_images}(d) can significantly reduce image noise and aliasing artifacts without degrading the diagnostic values. Despite of the clear benefits of the JENG algorithm, the current implementation of JENG does not correct the beam hardening artifacts. Therefore, mild beam hardening artifacts from the cancer nodule can be seen near the upper left lobe.

Fig.~\ref{Fig:coronal_clinical_images} is an example cross-plane image from coronal view. Fig.~\ref{Fig:coronal_clinical_images}(a) is reconstructed by the clinical standard hybrid IR and Fig.~\ref{Fig:coronal_clinical_images}(b) is reconstructed by JENG at a comparable resolution. A major advantage for JENG is its better detectability of small bone openings near the shoulder, better aliasing artifact reduction and more effective image denoising. In addition, JENG provides a much more uniform image texture for the soft tissues. In comparison, the clinical standard hybrid IR in Fig.~\ref{Fig:coronal_clinical_images}(a) has more difficulty to detect small bone openings and shows an inferior capability to suppress aliasing artifacts and image noise.

\begin{figure}
\begin{tikzpicture}[spy using outlines={circle,red,magnification=1.5,size=3.6cm, connect spies}]
\node {\includegraphics[width=.49\linewidth]{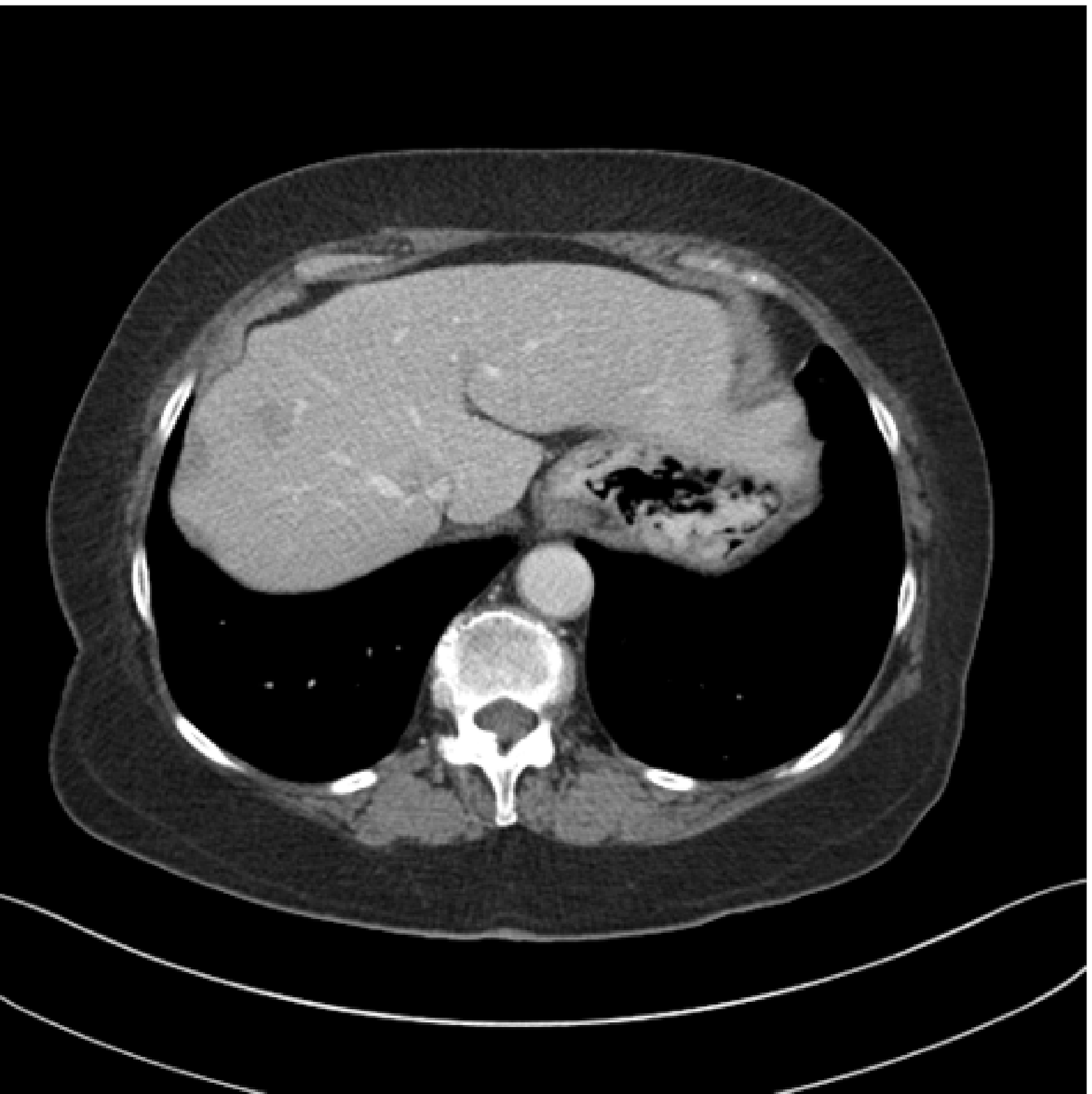}};
\spy on (-1.6,1.0) in node [left] at (-0.45,-2.75);
\node at (1.0,-4.4) {(a) Clinical Standard IR};
\end{tikzpicture}
\begin{tikzpicture}[spy using outlines={circle,red,magnification=1.5,size=3.6cm, connect spies}]
\node {\includegraphics[width=.49\linewidth]{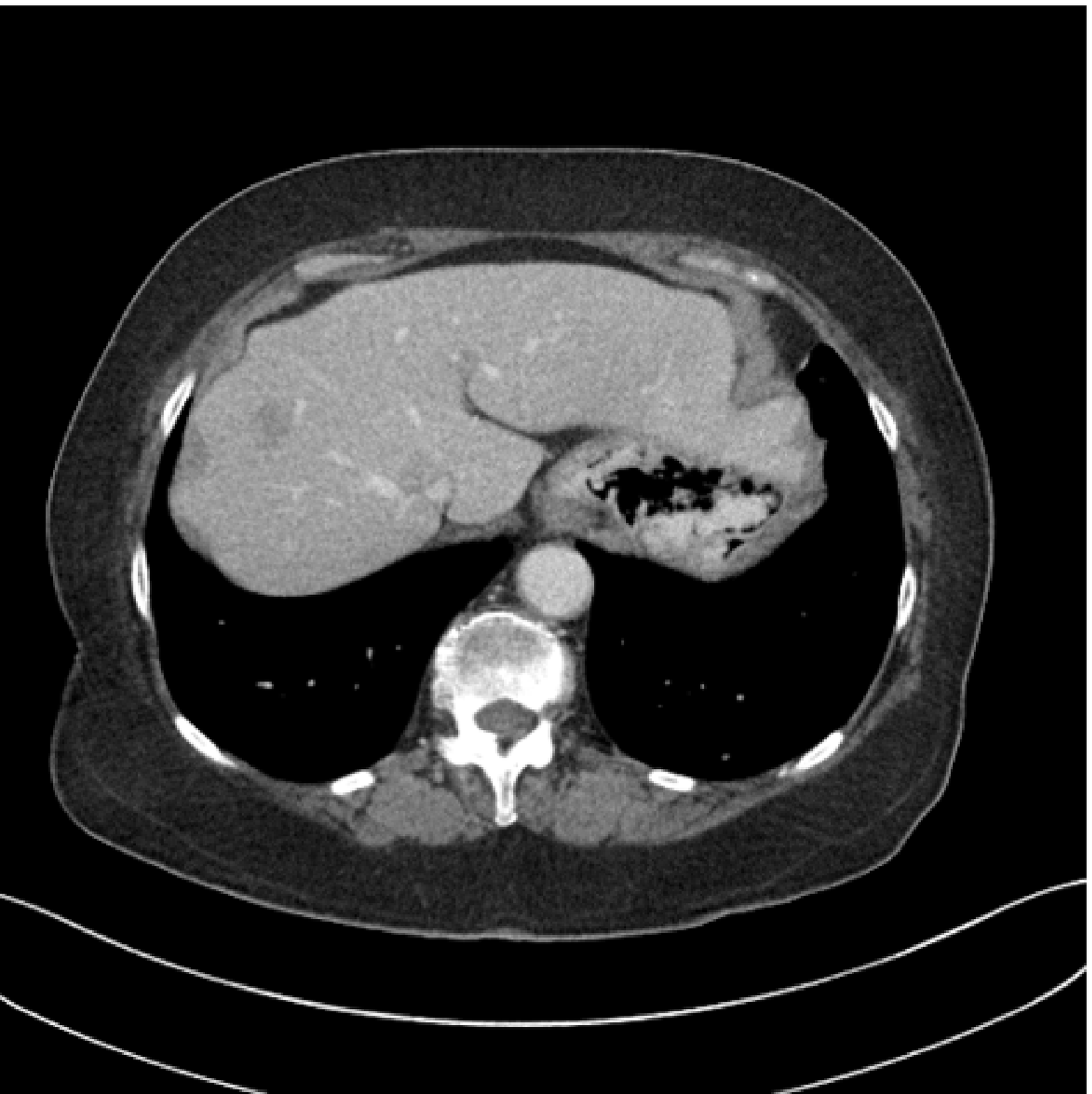}};
\spy on (-1.6,1.0) in node [left] at (-0.45,-2.75);
\node at (1.0,-4.4) {(b) JENG};
\end{tikzpicture}
\caption{An example liver image with a magnified sub-figure showing low-contrast cysts in the liver. (a) The clinical standard hybrid IR in soft tissue window. (b) JENG at a marginally better low-contrast resolution but with reduced image noise and artifacts.}
\label{fig:low_contrast}
  \vspace{1em}
\end{figure}

One concern people might have for JENG is that its better capability of artifacts removal and denoising might lead to a worse low-contrast lesion detectability. To address this concern and show that JENG does not have a compromised low-contrast lesion detectability, Fig.~\ref{fig:low_contrast} shows a liver image from a CT abdominal scan with magnified sub-figures for low-contrast liver cysts with Fig.~\ref{fig:low_contrast}(a) for clinical standard hybrid IR and Fig.~\ref{fig:low_contrast}(b) for JENG. Notice that the JENG image has less image noise and fewer artifacts than the clinical standard hybrid IR, despite that JENG also has a marginally better low-contrast detectability than the clinical standard with liver cysts clearly shown in the magnified sub-figure. In addition, the image texture in the liver is more uniform in JENG than the clinical standard hybrid IR. 

\section{Discussion and Conclusion}

Although an important feature for DS-FFS CT is to minimize scan time by enabling a high helical pitch while simultaneously reducing undersampling artifacts, it is not easy to fully leverage the acquisition physics and geometry of the scanner and the clinical standard methods have limited success to reduce undersampling artifacts. In addition, the projection data interpolation, rebinning and completion may also cause a compromised spatial resolution.  
In response, we presented the first physics-based iterative reconstruction algorithm for DS-FFS CT and used the native cone-beam geometry and the precise dual source helical trajectory to achieve higher spatial resolution and eliminate undersampling artifacts. Experimental results on phantom and clinical datasets show that our new algorithm, JENG, has a task-based MTF much higher than the clinical standard hybrid IR method while significantly reducing undersampling artifacts.

With a higher spatial resolution, radiologist can potentially use the JENG algorithm to better distinguish different objects or tissues located within a small proximity to each other. In the example of thoracic CT scans, JENG's improved image spatial resolution and contrast can lead to more accurate imaging on small indeterminate lung nodules and pulmonary emboli caused by intravascular disease.
In addition, given that undersampling artifacts are quantization artifacts caused by insufficient projections,  JENG's capability to reduce undersampling artifacts allows us to acquire fewer projections without compromising image diagnostic values, and thereby lower radiation doses received by patients.

Although JENG has the benefits discussed above, the NPS analysis shows that the existing implementation for JENG has an insufficient denoising capability at very low spatial frequency. Since the prior model for JENG is a Generalized Markov Random Field and is a low pass filter, JENG is successful in denoising high frequency content, but has limited success at very low frequency.  The relatively high NPS at very low spatial frequency ($<$0.1 mm$^{-1}$), however, is unlikely to impair diagnostic values. Since image signals often have substantial energy at low frequencies but have much less energy at high frequencies, the Signal Noise Ratio (SNR) is high at low frequencies and image signals are readily apparent despite the noise. In contrast, the SNR is lower at high frequencies and a prior model with a good high frequency denoising, such as the Markov Random Field prior model, can significantly improve image quality. One adverse effect from a low pass filtering prior model is that the noise texture of JENG can appear different from that of FBP, and can negatively influence a radiologist's image perception if he is not used to the noise texture of JENG.

Another drawback in this paper's technical contribution is the lack of modeling on focal spot size and this paper assumes that the focal spot is a sizeless point. For most CT scans that require a high spatial resolution, the focal spot size is often less than 1 mm. In these applications, this paper's focal spot modeling as a sizeless point is a good assumption for high image quality. In a few other applications that require a large focal spot and reconstructions might benefit more from a precise focal spot size modeling, the sizeless point assumption can potentially lead to a loss of spatial resolution. 

\section*{Acknowledgment}
This investigation was supported by DOE grant (No. DEAC02-05CH11231), NSF grant (No. CCF-1763896) and a Young Investigator Award from the Society for Pediatric Radiology.

\section*{Data Availability Statement}
The data that support the findings of this study are available on request from the corresponding author. The data are not publicly available due to privacy or ethical restrictions.


\section*{References}
\addcontentsline{toc}{section}{\numberline{}References}
\vspace*{-20mm}










\bibliographystyle{./medphy.bst}    


\end{document}